\pdfoutput=1

\documentclass[11pt,twoside,a4paper,cmspaper,final,collab]{cms-tdr}
\def\svnVersion{272422}\def\cmsCernNo{2013-037}\def\cmsCernDate{\today}\def\cmsMessage{Published in Physics Letters B as \href{http://dx.doi.org/10.1016/j.physletb.2014.11.059}{\doi{10.1016/j.physletb.2014.11.059.}}}

\begin{document}\cmsNoteHeader{SMP-13-005}

\hyphenation{had-ron-i-za-tion}
\hyphenation{cal-or-i-me-ter}
\hyphenation{de-vices}
\RCS$Revision: 272194 $
\RCS$HeadURL: svn+ssh://svn.cern.ch/reps/tdr2/papers/SMP-13-005/trunk/SMP-13-005.tex $
\RCS$Id: SMP-13-005.tex 272194 2014-12-18 11:30:18Z asavin $
\newlength\cmsFigWidth
\ifthenelse{\boolean{cms@external}}{\setlength\cmsFigWidth{0.48\textwidth}}{\setlength\cmsFigWidth{0.65\textwidth}}
\ifthenelse{\boolean{cms@external}}{\providecommand{\cmsLeft}{top}}{\providecommand{\cmsLeft}{left}}
\ifthenelse{\boolean{cms@external}}{\providecommand{\cmsRight}{bottom}}{\providecommand{\cmsRight}{right}}
\newcommand{\tauh}{\ensuremath{\Pgt_\mathrm{h}}}
\newcommand{\taumu}{\ensuremath{\Pgt_{\Pgm}}}
\newcommand{\taue}{\ensuremath{\Pgt_{\Pe}}}
\newcommand{\taul}{\ensuremath{\Pgt_{\ell}}}
\newcommand{\taus}{\ensuremath{\ell\ell\Pgt\Pgt}}
\providecommand{\PZ}{\cPZ\xspace}
\newcommand{\ZZ}{\cPZ\cPZ\xspace}
\newcommand{\WW}{\PWp\PWm\xspace}
\newcommand{\WZ}{\PW\cPZ\xspace}
\newcommand{\pp}{\Pp\Pp\xspace}
\newcommand{\Wjets}{\ensuremath{\PW+\text{jets}}\xspace}
\hyphenation{ATGCs}
\newcommand{\fourl}{\ensuremath{\ell^{+}_\mathrm{i} \ell^{-}_\mathrm{i} \ell^{+}_\mathrm{j} \ell^{-}_\mathrm{j}}}
\newcommand{\ZZfourl}{\ensuremath{{\PZ\PZ} \to 2\ell 2\ell}}
\newcommand{\ZZtwoltwotau}{\ensuremath{{\PZ\PZ} \to 2l2\tau}}
\providecommand{\mH}{\ensuremath{m_{\PH}\xspace}}
\providecommand{\vecEtm}{\ensuremath{\vec{E}_\mathrm{T}^{\mathrm{miss}}}\xspace}
\newcommand{\dytt}{\ensuremath{\cPZ/\Pgg^*\to\tau^+\tau^-}}
\newcommand{\dyll}{\ensuremath{\cPZ/\Pgg^*\to \ell^+\ell^-}}
\newcommand{\dyee}{\ensuremath{\cPZ/\Pgg^*\to\Pep\Pem}}
\newcommand{\dymm}{\ensuremath{\cPZ/\Pgg^*\to\Pgmp\Pgmm}}
\newcommand{\tw}{\cPqt\PW}
\newcommand{\wgamma}{\PW\Pgg}
\renewcommand{\syst}{\ensuremath{\,\mathrm{(syst)}}\xspace}
\newcommand{\theo}{\ensuremath{\,\mathrm{(theo)}}\xspace}
\renewcommand{\lumi}{\ensuremath{\,\mathrm{(lumi)}}\xspace}
\cmsNoteHeader{SMP-13-005}
\title{Measurement of the $\Pp\Pp \to \cPZ\cPZ$ production cross section  and constraints on
anomalous triple gauge couplings in four-lepton final states at $\sqrt{s} = 8\TeV$}

\author{CMS Collaboration}

\date{\today}

\abstract{
A measurement of the inclusive $\PZ\PZ$ production cross section and constraints on
anomalous triple gauge couplings in proton-proton collisions at $\sqrt{s} = 8\TeV$ are presented.
The analysis is based on a data sample, 
corresponding to an integrated luminosity of 19.6\fbinv,
collected with the CMS experiment at the LHC.
The measurements are performed in the leptonic decay modes $\cPZ\cPZ \to \ell\ell\ell'\ell'$, where
$\ell = \Pe, \Pgm$ and $\ell' = \Pe, \Pgm, \Pgt$.
The measured total cross section
$\sigma ( \pp \to \PZ\PZ) = 7.7\, \pm 0.5\stat\, ^{+0.5} _{-0.4}\syst \pm 0.4\theo \pm 0.2\lum\unit{pb}$,
for both $\cPZ$ bosons produced in the mass range $60 < m_{\cPZ} < 120\GeV$,
is consistent with standard model predictions. Differential cross sections are
measured and well described by the theoretical  predictions.
The invariant mass distribution of the four-lepton system is used to set limits on anomalous $\cPZ\cPZ\cPZ$ and $\cPZ\cPZ\gamma$ couplings at the 95\% confidence level: $-0.004<f_4^\cPZ<0.004$, $-0.004<f_5^\cPZ<0.004$, $-0.005<f_4^{\gamma}<0.005$, and $-0.005<f_5^{\gamma}<0.005$.
}

\hypersetup{%
pdfauthor={CMS Collaboration},%
pdftitle={Measurement of the pp to ZZ production cross section and constraints on anomalous triple gauge couplings in four-lepton final states at sqrt(s) = 8 TeV},%
pdfsubject={CMS},%
pdfkeywords={CMS, physics, Electroweak}}

\maketitle

\section{Introduction}

The study of diboson production in proton-proton collisions provides an important
test of the electroweak sector of the standard model (SM),
especially the non-Abelian structure of the SM Lagrangian.
In the SM, $\cPZ\cPZ$ production proceeds mainly through quark-antiquark
$t$- and $u$-channel scattering
diagrams. At high-order calculations in QCD, gluon-gluon fusion also contributes via
box diagrams with quark loops. There are no tree-level contributions to
$\cPZ\cPZ$ production from triple gauge boson vertices in the SM.
Anomalous triple gauge couplings (ATGC)
$\PZ\PZ\PZ$ and $\PZ\PZ\gamma$ are introduced using an effective Lagrangian
following Ref.~\cite{Hagiwara}.
In this parametrization, two $\PZ\PZ\PZ$ and two $\PZ\PZ\gamma$ couplings are allowed by electromagnetic gauge
invariance and Lorentz invariance for on-shell $\PZ$ bosons
and are parametrized by two
CP-violating ($f_4^{\mathrm{V}}$) and two CP-conserving ($f_5^{\mathrm{V}}$)
parameters,
where ${\mathrm{V}} = (\PZ, \gamma)$.
Nonzero ATGC values could be induced by new physics models such as
supersymmetry~\cite{Gounaris:2000tb}.

Previous measurements of the inclusive ZZ cross section by the CMS Collaboration
at the LHC were performed in the
$\cPZ\cPZ \to \ell\ell\ell'\ell'$ decay channels, where
$\ell = \Pe, \Pgm$ and $\ell' = \Pe, \Pgm, \Pgt$,
with the data corresponding to an integrated luminosity of
5.1 (5.0)\fbinv at $\sqrt{s} = 7(8)\TeV$~\cite{SMP-12-007,SMP-12-024}.
The measured total cross section, $\sigma ( \pp \to \cPZ\cPZ)$,
is
$6.24\, ^{+0.86}_{-0.80}\stat\, ^{+0.41}_{-0.32}\syst \pm 0.14\lum \unit{pb}$ at $\sqrt{s}=7\TeV$
and $8.4 \pm 1.0\stat \pm 0.7\syst \pm 0.4\lum$  $\unit{pb}$ at $\sqrt{s} = 8\TeV$
for both $\cPZ$ bosons in the mass range $60 < m_\cPZ < 120\GeV$.
The ATLAS Collaboration  measured
a total cross section of
$6.7 \pm 0.7\stat\, ^{+0.4}_{-0.3}\syst \pm 0.3\lum\unit{pb}$~\cite{ATLASnewXS}
using $\cPZ\cPZ \to \ell\ell\ell\ell$ and $\cPZ\cPZ \to \ell\ell\nu\nu$ final states
with a data sample corresponding to an integrated luminosity of $4.6\fbinv$ at
$\sqrt{s} = 7\TeV$ and $66 < m_\cPZ < 116\GeV$.
Measurements of the $\cPZ\cPZ$ cross sections performed at
the Tevatron are summarized in Refs.~\cite{CDFZZxs,D0ZZxs}.
All measurements are found to agree with the corresponding SM predictions.

Limits on $\cPZ\cPZ\PZ$ and $\cPZ\cPZ\gamma$ ATGCs
were set by CMS using the $7\TeV$ data sample:
$-0.011<f_4^\cPZ<0.012$, $-0.012<f_5^\cPZ<0.012$, $-0.013<f_4^{\gamma}<0.015$, and
$-0.014<f_5^{\gamma}<0.014$ at 95\% confidence level (CL)~\cite{SMP-12-007}.
Similar limits were obtained by ATLAS~\cite{ATLASnewXS}.

In this analysis, which is based on the full 2012 data set and
corresponds to an integrated luminosity of 19.6\fbinv,
results are presented for the $\ZZ$ inclusive and differential
cross sections 
as well as limits for the
$\cPZ\cPZ\cPZ$ and $\cPZ\cPZ\gamma$ ATGCs.
The cross sections are measured for both $\cPZ$ bosons in the mass 
range $60 < m_\cPZ < 120\GeV$;
contributions from virtual photon exchange are included.

\section{The CMS detector and simulation}

The CMS detector is described in detail elsewhere~\cite{CMSExperiment};
the
key components for this analysis are summarized here.
The CMS experiment uses a right-handed coordinate system, with the
origin at the nominal interaction point, the $x$ axis
pointing to the center of the LHC ring, the $y$ axis pointing
up (perpendicular to the plane of the LHC ring), and the $z$ axis
along the counterclockwise-beam direction. The polar angle
$\theta$ is measured from the positive $z$ axis and the
azimuthal angle $\phi$ is measured in the $x$-$y$ plane.
The magnitude of the transverse
momentum is $\pt = \sqrt{\smash[b]{p_x^2 + p_y^2}}$.
A superconducting solenoid is located in the
central region of the CMS detector, providing an axial magnetic
field of 3.8\unit{T} parallel to the beam direction.
A silicon pixel and strip
tracker, a crystal electromagnetic calorimeter (ECAL), and a brass and scintillator hadron
calorimeter are located within the solenoid and cover the absolute pseudorapidity
range
 $\abs{ \eta } < 3.0$,
where pseudorapidity
is defined as $\eta=-\ln[\tan{(\theta/2)}]$.
The ECAL barrel region (EB) covers $\abs{\eta} < 1.479$ and two endcap regions (EE) cover
$1.479 < \abs{\eta} < 3.0$.
A quartz-fiber
Cherenkov calorimeter extends the coverage up to $\abs{\eta} <$ 5.0.
Gas ionization muon detectors are embedded
in the steel flux-return yoke outside the solenoid.
A first level of the CMS trigger system, composed of custom
hardware processors, is designed to select events of interest
in less than 4\mus using information from the calorimeters and muon
detectors.
A high-level-trigger processor farm reduces the event rate from 100\unit{kHz}
delivered by the first level trigger to a few hundred hertz.

Several Monte Carlo (MC) event generators are used to simulate the signal and
background contributions.
The $\Pq\Paq \to \ZZ$ process is generated at
next-to-leading order (NLO) with
\POWHEG2.0~\cite{Alioli:2008gx,Nason:2004rx,Frixione:2007vw}
or at leading-order (LO) with \SHERPA~\cite{Sherpa}.
The $\Pg\Pg \to \ZZ$ process
is simulated
with {\textsc{gg2zz}}~\cite{Binoth:2008pr} at LO.
Other diboson processes ($\PW\cPZ$, $\cPZ\gamma$ )
and the $\cPZ$+jets samples
are generated at LO with \MADGRAPH5~\cite{Madgraph5}.
Events from $\ttbar$ production are generated at NLO with
\POWHEG. The
\PYTHIA6.4~\cite{Sjostrand:2006za} package is used for parton showering, hadronization, and
the underlying event simulation.
The default set of parton distribution functions
(PDF) used for LO generators is CTEQ6L~\cite{CTEQ66}, whereas
CT10~\cite{ct10} is used for NLO generators.
The $\PZ\PZ$ yields from simulation are scaled according to the
theoretical cross sections
calculated with \MCFM 6.0~\cite{MCFM} at
NLO for $\Pq\Pq \to \PZ\PZ$ and at
LO for $\Pg\Pg \to \PZ\PZ$
with
the MSTW2008 PDF~\cite{Martin:2009iq}
with renormalization and factorization scales set to $\mu_{\mathrm{R}} = \mu_{\mathrm{F}} = 91.2\GeV$.
The $\Pgt$-lepton decays are simulated with {\TAUOLA}~\cite{Jadach:1993hs}.
For all processes, the detector response is simulated using a detailed
description of the CMS detector based on the \GEANTfour~
package~\cite{GEANT}, and event reconstruction is performed with
the same algorithms that are used for data.
The simulated samples include multiple interactions per bunch crossing (pileup), such
that the pileup distribution matches that of data,
with an average value of about 21 interactions per bunch crossing.

\section{Event reconstruction}
\label{sec:eventreconstruction}

A complete reconstruction of the individual particles emerging from each collision
event  is obtained via a particle-flow (PF)
technique~\cite{CMS-PAS-PFT-09-001,CMS-PAS-PFT-10-001}, which
uses the information from all CMS sub-detectors to identify and reconstruct individual particles
in the collision event. The particles  are classified into
mutually exclusive categories: charged hadrons, neutral hadrons, photons, muons, and electrons.

Electrons are reconstructed within the geometrical acceptance, $\abs{\eta^{\Pe}} < 2.5$, and
for transverse momentum $\pt^{\Pe} > 7\GeV$.
The reconstruction combines the information from clusters of energy deposits in the ECAL and the
trajectory in the 
tracker~\cite{Baffioni:2006cd}.
Particle trajectories in the tracker volume are reconstructed using a  modeling of the electron
energy loss and fitted with a Gaussian sum filter~\cite{gsf}.
The contribution of the ECAL energy deposits to the electron transverse momentum measurement and its uncertainty
are determined via a multivariate regression approach.
Electron identification relies on a multivariate technique that combines observables sensitive to the
amount of  bremsstrahlung along the electron trajectory, the geometrical and momentum matching
between the electron trajectory and associated clusters, as well as shower shape observables.

Muons are reconstructed within $\abs{\eta^{\Pgm}} < 2.4$ and for $\pt^{\Pgm} > 5\GeV$~\cite{Chatrchyan:2012xi}.
The reconstruction combines information from both the silicon tracker and the
muon detectors.
The PF muons are selected from among the reconstructed muon track candidates
by applying requirements on the track components in the muon system
and matching with minimum ionizing particle energy deposits in the
calorimeters.

For $\Pgt$ leptons, two principal decay modes are distinguished: a leptonic mode, $\taul$, with a final state including
either an
electron or a muon, and a hadronic mode,
$\tauh$, with a final state including hadrons.
The PF particles are used to reconstruct $\tauh$ with the ``hadron-plus-strip''
 algorithm~\cite{Chatrchyan:2011xq}, which
 optimizes the reconstruction and identification of specific
$\tauh$ decay modes.
The $\pi^0$ components of the $\tauh$ decays are first reconstructed and then combined with charged
hadrons to reconstruct the $\tauh$ decay modes. Cases where
$\tauh$ includes three charged hadrons are also included.
The missing transverse energy  that is associated with neutrinos from $\tau$ decays is ignored in the reconstruction.
The $\tauh$ candidates in this analysis are required to have $\abs{\eta^{\tauh}} < 2.3$ and $\pt^{\tauh} > 20\GeV$.

The isolation of individual electrons or muons is measured relative to their transverse momentum $\pt^{\ell}$,
by summing over the transverse momenta of charged hadrons and neutral particles in a cone with
$\Delta R = \sqrt{\smash[b]{(\Delta \eta)^{2} + (\Delta \phi)^{2}}} = 0.4$ around the lepton direction
at the interaction vertex:
\ifthenelse{\boolean{cms@external}}{
\begin{multline}
R_\text{Iso}^{\ell} = \Bigg( \sum  \pt^\text{charged}\\
+ \text{MAX}\Big[ 0, \sum \pt^\text{neutral}
+  \sum \pt^{\gamma} - \rho \times A_\text{eff}  \Big] \Bigg) \Big/  \pt^{\ell}.
\end{multline}
}{
\begin{equation}
R_\text{Iso}^{\ell} = \left( \sum  \pt^\text{charged} + \text{MAX}\left[ 0, \sum \pt^\text{neutral}
                                 +  \sum \pt^{\gamma} - \rho \times A_\text{eff}  \right] \right) /  \pt^{\ell}.
\end{equation}
}
The $\sum  \pt^\text{charged}$ is the scalar sum of the transverse momenta of charged hadrons
originating from the primary vertex.
The primary vertex is chosen as the vertex with the highest sum of $\pt^2$ of
its constituent tracks.
The $\sum \pt^\text{neutral}$ and $\sum \pt^{\gamma}$ are the
scalar sums of the transverse momenta for neutral hadrons and photons, respectively.
The average transverse-momentum flow density $\rho$ is calculated in each event using a ``jet area''~\cite{Cacciari:2007fd}, where
 $\rho$ is
defined as the median of the $\pt^\text{jet}/A_\text{jet}$ distribution for all pileup jets in the event.
The effective area $ A_\text{eff}$ is the geometric area of the isolation cone times an $\eta$-dependent correction factor
that accounts for the
residual dependence of the isolation on pileup.
Electrons and muons are considered isolated
if $  R_\text{Iso}^{\ell} < 0.4 $.
Allowing $\Pgt$ leptons in the final state increases the background
contamination, therefore
tighter isolation requirements are imposed for electrons and muons in $\cPZ\cPZ \to \taus$
decays:
$ R_\text{Iso}^{\ell} < 0.25 $ for
$ \cPZ \to \Pgt_{\ell}^+ \Pgt_{\ell}^-$, and
$ R_\text{Iso}^{\Pe} < 0.1 $ for
$ \cPZ \to \Pgt_{\Pe} \tauh$, and
$ R_\text{Iso}^{\Pgm} < 0.15$ for
$\Pgt_{\mu} \tauh$.

The isolation of the $\tauh$ is calculated as the scalar sum of the transverse momenta of the
charged hadrons and neutral particles in a cone of $\Delta R = 0.5$
around the $\tauh$ direction reconstructed at the interaction vertex.
The $\tauh$ isolation includes a correction for pileup effects,
which is based on the scalar sum of transverse momenta of charged particles
 not associated with the primary vertex in a cone of $\Delta R = 0.8$ about
the $\tauh$ candidate direction ($\pt^\mathrm{PU}$).
The isolation variable is defined as:
\ifthenelse{\boolean{cms@external}}{
\begin{multline}
I^\mathrm{PF} =  \Bigg ( \sum \pt^\text{charged}\\+ \text{MAX}\Big[ 0, \sum  \pt^\text{neutral}
+ \sum \pt^{\gamma}- f \times \pt^\mathrm{PU} \Big] \Bigg),
\end{multline}
}{
\begin{equation}
I^\mathrm{PF} =  \left ( \sum \pt^\text{charged}+ \text{MAX}\left[ 0, \sum  \pt^\text{neutral} + \sum \pt^{\gamma}- f \times \pt^\mathrm{PU} \right] \right),
\end{equation}
}
where the scale factor of $f = 0.0729$, which is
used in estimating the contribution to the isolation sum
from neutral hadrons and photons, accounts for the difference in the neutral and charged contributions and in the cone sizes. Two standard
working points are defined based on the value of the isolation sum corrected for the pileup contribution:
$I^{\mathrm{PF}} < 1\, (8)\GeV$
for final states including one (two) $\tauh$ candidates.

The electron and muon pairs from $\cPZ$-boson decays are required to originate from the primary vertex.
This is ensured by demanding that the significance of the three-dimensional impact parameter relative
to the event
vertex, $\mathrm{SIP_{3D}}$, satisfies $\mathrm{SIP_{3D}} = \abs{ \frac{\mathrm{IP}}{\sigma_\mathrm{IP}} } < 4$
for each lepton. The $\mathrm{IP}$  is the distance
of closest approach of the lepton track to the primary vertex
and
$\sigma_\mathrm{IP}$ is its associated uncertainty.

The combined efficiencies of reconstruction, identification, and isolation of
primary electrons or muons are measured in data using a ``tag-and-probe''
technique~\cite{CMS:2011aa} applied to an inclusive sample of $\cPZ$ events.
The measurements are performed in bins of $\pt^{\ell} $ and $\abs{\eta^{\ell}}$.
The efficiency for selecting electrons in the ECAL barrel(endcaps) is
about  70\%(60\%) for $7 < \pt^{\Pe} < 10\GeV$,  85\%(77\%) at $\pt^{\Pe} \simeq 10\GeV$,
and 95\%(89\%) for $\pt^{\Pe} \geq 20\GeV$.
It is about 85\% in the transition region
between
the ECAL barrel and endcaps
($1.44 < \abs{\eta} < 1.57$),
averaging over the whole $\pt$ range.
The muons are reconstructed and identified with an efficiency greater than ${\sim}98\%$ in the
full $\abs{\eta^{\Pgm}} < 2.4$ range.
The $\tauh$ reconstruction efficiency is approximately 50\%~\cite{Chatrchyan:2011xq}.

Final-state radiation (FSR) may affect the measured four-momentum of the leptons if it is
not properly included in the reconstruction. For electrons,
a significant portion of the FSR photons is
included in the reconstructed energy because of the size of the electromagnetic clusters, but for muons additional treatment of the
FSR photons is important.
All photons reconstructed within $\abs{\eta^{\Pgm}} < 2.4$ are considered as possible FSR candidates if
they  have a transverse momentum $\pt^\gamma > 2(4)\GeV $  and are found
within $ \Delta R < 0.07 (0.07 < \Delta R < 0.5) $ from the closest selected lepton candidate
and are isolated.
The photon isolation observable $R_\text{Iso}^{\gamma}$ is the sum
of the transverse momenta
of charged hadrons, neutral hadrons, and photons
in a cone of $\Delta R = 0.3$ around the candidate photon
direction, divided by $\pt^\gamma$.
Isolated photons must satisfy $R_\text{Iso}^{\gamma} < 1$.
The recovered FSR photon is included in the lepton four-momentum and the lepton isolation is then recalculated without it.

The performance of the FSR selection algorithm has been determined using
simulated samples, and the rate is verified with the $\cPZ$ and $\cPZ\cPZ$ events in data.
The photons within the acceptance for the FSR selection are reconstructed
with an efficiency of about 50\% and with a mean purity of 80\%.
The FSR photons are recovered in 0.5(5)\% of  inclusive $\cPZ$  events with
electron (muon)  pairs.

\section{Event selection}

The data sample used in this analysis is selected by the trigger system,
which requires the presence of a
pair of electrons or  muons, or a triplet of electrons.
Triggers requiring an electron and a muon are also used.
For the double-lepton triggers, the highest $\pt$ and the second-highest $\pt$ leptons are required to have $\pt$ greater than 17 and 8\GeV, respectively,
while for the triple-electron trigger the thresholds are 15, 8, and 5\GeV.
The trigger efficiency for $\cPZ\cPZ$ events within the acceptance of this analysis is greater than 98\%. The use of the triple-electron trigger with a looser $\pt$ requirement helps to recover 1-2\% of the signal events, while for muons such contribution was found to be negligible.

In selected $\cPZ\cPZ$ events, the $\cPZ$ candidate with the mass closest
to the $\cPZ$-boson mass is denoted $\cPZ_1$ and the other one, $\cPZ_2$. The selection is designed
 to give mutually exclusive sets of signal candidates first selecting $\cPZ\cPZ$ decays to $4\Pe$, $4\Pgm$, and $2\Pe 2\Pgm$, in the following denoted $\ell\ell\ell''\ell''$; these events are not considered in $\PZ\PZ\to \taus$ channel.
The leptons are identified and isolated as described in Section~\ref{sec:eventreconstruction}.
When building the $\cPZ$ candidates, the FSR photons are kept if
$\abs {m_{\ell\ell\gamma} - m_\cPZ} < \abs {m_{\ell\ell} - m_\cPZ}$ and
$m_{\ell\ell\gamma} < 100\GeV$.
In the following, the presence of the photons in the $\ell\ell\ell''\ell''$ kinematics is implicit.
The leptons constituting a $\cPZ$ candidate are required to be the same
flavor and  to have opposite charges ($\ell^+\ell^-$).
The pair is
retained if it satisfies $60 < m_{\cPZ} < 120\GeV$.
If more than one ${\cPZ_2}$ candidate satisfies all criteria, the ambiguity
is resolved by choosing the pair of leptons with the highest scalar sum of $\pt$.
Among the four selected leptons forming the  $\cPZ_1$ and the $\cPZ_2$,
at least one should have $\pt > 20\GeV$ and another one should have
 $\pt > 10\GeV$.
These \pt thresholds ensure that the selected events have leptons
with $\pt$ values on the high-efficiency plateau for the trigger.

For the $\taus$ final state, events are required to have one
$\cPZ_1 \to \ell^+\ell^-$ candidate with $\pt > 20\GeV$ for one of the leptons
and $\pt > 10\GeV$ for the other lepton,
and a  $\cPZ_2 \to \Pgt^+\Pgt^-$, with $\Pgt$ decaying into $\taue$, $\taumu$, or $\tauh$.
The leptons from the $\Pgt_\ell$  decays are required to have $\pt^{\ell} > 10\GeV$.
The $\tauh$ candidates are required to have $\pt^{\tauh} > 20\GeV$.
The FSR recovery is not applied to the $\taus$ final states, since it does not improve the mass reconstruction.
The invariant mass of the reconstructed $\cPZ_1$ is required to satisfy
$ 60 < m_{\ell\ell} < 120\GeV$, and that of the $\cPZ_2$ to satisfy
$m_{\text{min}} < m_{\Pgt\Pgt} < 90\GeV$, where $m_{\text{min}} = 20\GeV$ for $\cPZ_2 \to \Pgt_{\Pe}\Pgt_{\mu}$
final states
and 30\GeV for all others.

\section{Background estimation}

The lepton identification and isolation requirements described in
Section~\ref{sec:eventreconstruction}
significantly suppress all background contributions, and the remnant portion of
them arise mainly from the $\PZ$ and $\PW\PZ$ production in association with
jets, as well as $\ttbar$.
In all these cases, a
jet or a non-prompt lepton is misidentified as an isolated  $\Pe$, $\Pgm$, $\tauh$, $\taue$, or $\taumu$.
Leptons produced in the decay of 
$\cPZ$ bosons are referred to as prompt leptons; 
leptons from e.g. heavy meson decays are non-prompt. The requirements to eliminate non-prompt leptons also remove hadrons that appear to be leptons.
 
To estimate the expected number of background events in the signal region,
control data samples
are defined for each lepton flavor combination $\ell'\ell'$.
The $\Pe$ and $\taue$, and $\Pgm$ and $\taumu$ are considered as different
flavors, since they originate from different particles.

The control data samples for the background estimate
are obtained by selecting events containing
$\cPZ_1$, which passes all selection requirements,
 and two additional lepton candidates $\ell'\ell'$.
The additional lepton pair must have opposite charge and matching
flavor ($\Pe^{\pm}\Pe^{\mp}, \Pgm^{\pm}\Pgm^{\mp}, \Pgt^{\pm}\Pgt^{\mp}$).
Control data samples enriched with
$\cPZ$+X events,
where X stands for $\bbbar$, $\ccbar$,
gluon, or light quark jets, are obtained by requiring that both
additional leptons pass
only relaxed identification criteria and are required to be not isolated.
By requiring one of the additional leptons to
pass the full selection requirements, one obtains data samples enriched
with $\PW\PZ$ events and significant number of $\ttbar$ events.
The expected number of background events in the
signal region for each flavor pair is obtained
by
scaling the number of observed $\cPZ_1+\ell'\ell'$ events
by the lepton misidentification probability and
combining the results for
$\cPZ$+X and
$\PW\PZ$, $\ttbar$ control regions together.
The procedure is identical for all lepton flavors.

The misidentification probability, \ie, the probability for a lepton candidate
that passes the relaxed requirements to pass the full selection,
is  measured separately for each flavor from a sample of
$\cPZ_1 + \ell_\text{candidate}$ events with a relaxed identification and
no isolation requirements on the $\ell_\text{candidate}$. The misidentification
probability for each lepton flavor is defined as the ratio of the number of leptons that
pass the final
isolation and identification requirements to the total number of leptons in
the sample.
It is measured in bins of lepton $\pt$ and $\eta$.
The contamination from $\PW\PZ$ events, which may lead to an overestimate of the
misidentification probability because of the presence of genuine isolated leptons,
is suppressed by requiring that the measured missing transverse energy is
less than 25\GeV.

The estimated background contributions
to the signal region are summarized in Table~\ref{table:results}.
The procedure
excludes a possible double counting due to $\cPZ$+X events that can be found in the 
$\PW\cPZ$ control region. A correction
for the small contribution of $\cPZ\cPZ$ events in the control region
is applied based on MC simulation.
The
predicted background yield has a small effect on
the $\ZZ$ cross section measurement in the $\ell\ell\ell''\ell''$ channels, but is
comparable to the signal yield for the case of $\taus$.

\section{Systematic uncertainties}

The systematic uncertainties for trigger efficiency (1.5\%) are evaluated from data.
The uncertainties arising from  lepton
identification and isolation are 1--2\% for muons and electrons, and 6--7\%
for $\tauh$.
The uncertainty in the LHC integrated luminosity of the data sample is 2.6\%~\cite{CMS-PAS-LUM-13-001}.

Theoretical uncertainties in the $\ZZ \to \ell\ell\ell''\ell''$ acceptance are evaluated using \MCFM
and by varying the renormalization and factorization scales, up and down, by a factor of two with respect to the default values
$\mu_{\mathrm{R}} = \mu_{\mathrm{F}} = m_\cPZ$. The variations in the acceptance are 0.1\% (NLO $\Pq\Paq \to \ZZ$) and
0.4\% ($\Pg\Pg \to \ZZ$), and can be neglected. Uncertainties related to the choice of the PDF and the strong coupling constant
$\alpha_s$ are evaluated following the PDF4LHC~\cite{Botje:2011sn} prescription and
using CT10, MSTW08, and NNPDF~\cite{nnpdf} PDF sets and found to be 4\% (NLO $\Pq\Paq \to \ZZ$) and 5\% ($\Pg\Pg \to \ZZ$).

The uncertainties in $\cPZ$+jets, $\PW\PZ$+jets, and $\ttbar$ yields reflect
the uncertainties in the measured values of the misidentification rates and the
limited statistics of the control regions in the data, and vary between 20\% and 70\%.

The uncertainty in the unfolding
procedure discussed in Section~\ref{sec:results}
 arises from differences between
\SHERPA and \POWHEG for the
unfolding factors (2--3\%),
from scale and PDF uncertainties (4--5\%),
and from experimental uncertainties (4--5\%).

\section{The \texorpdfstring{$\cPZ\cPZ$}{ZZ} cross section measurement}
\label{sec:results}

\begin{figure*}[htbp]
 \centering
\includegraphics[width=0.45\textwidth]{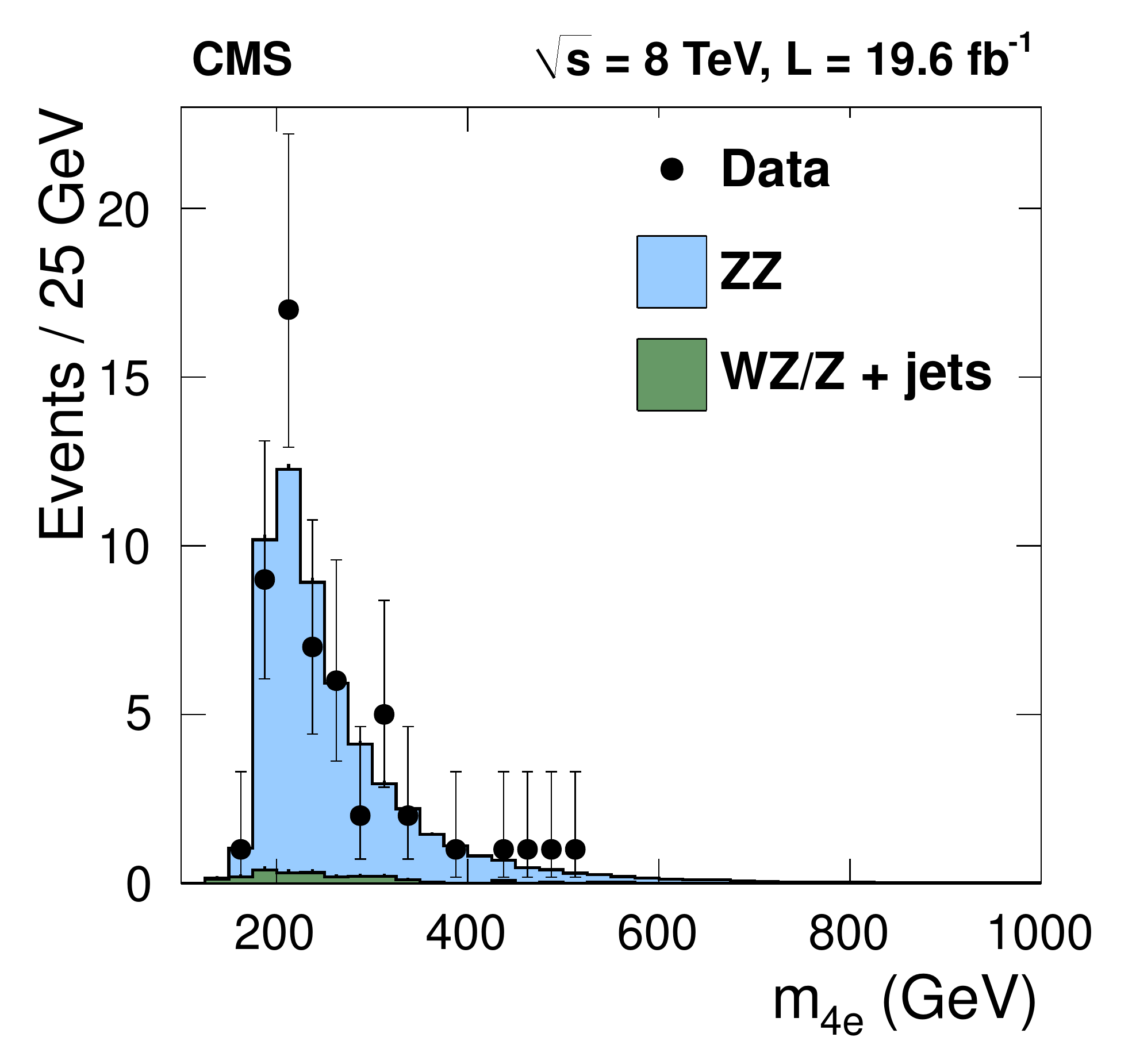}
\includegraphics[width=0.45\textwidth]{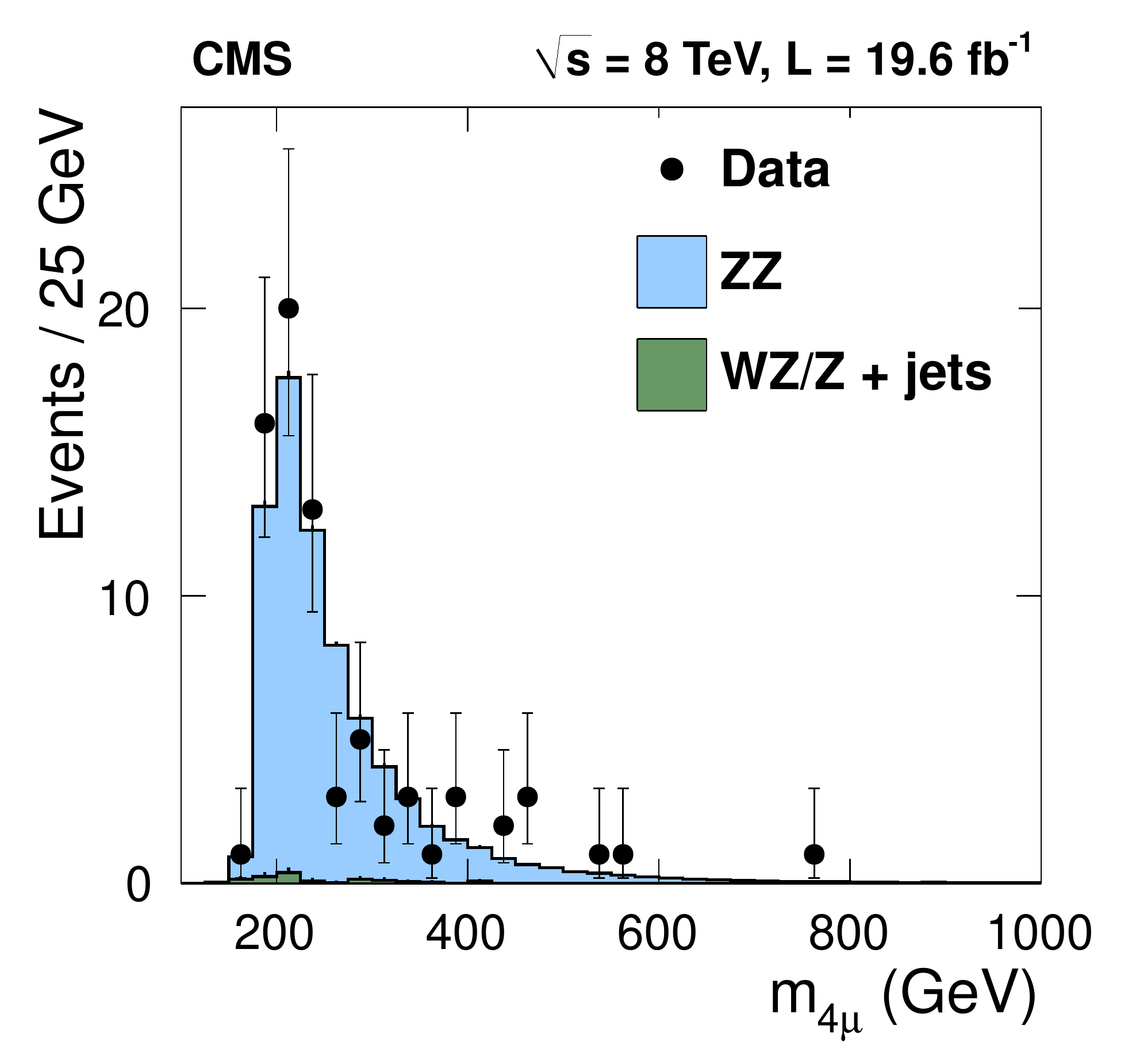}
\includegraphics[width=0.45\textwidth]{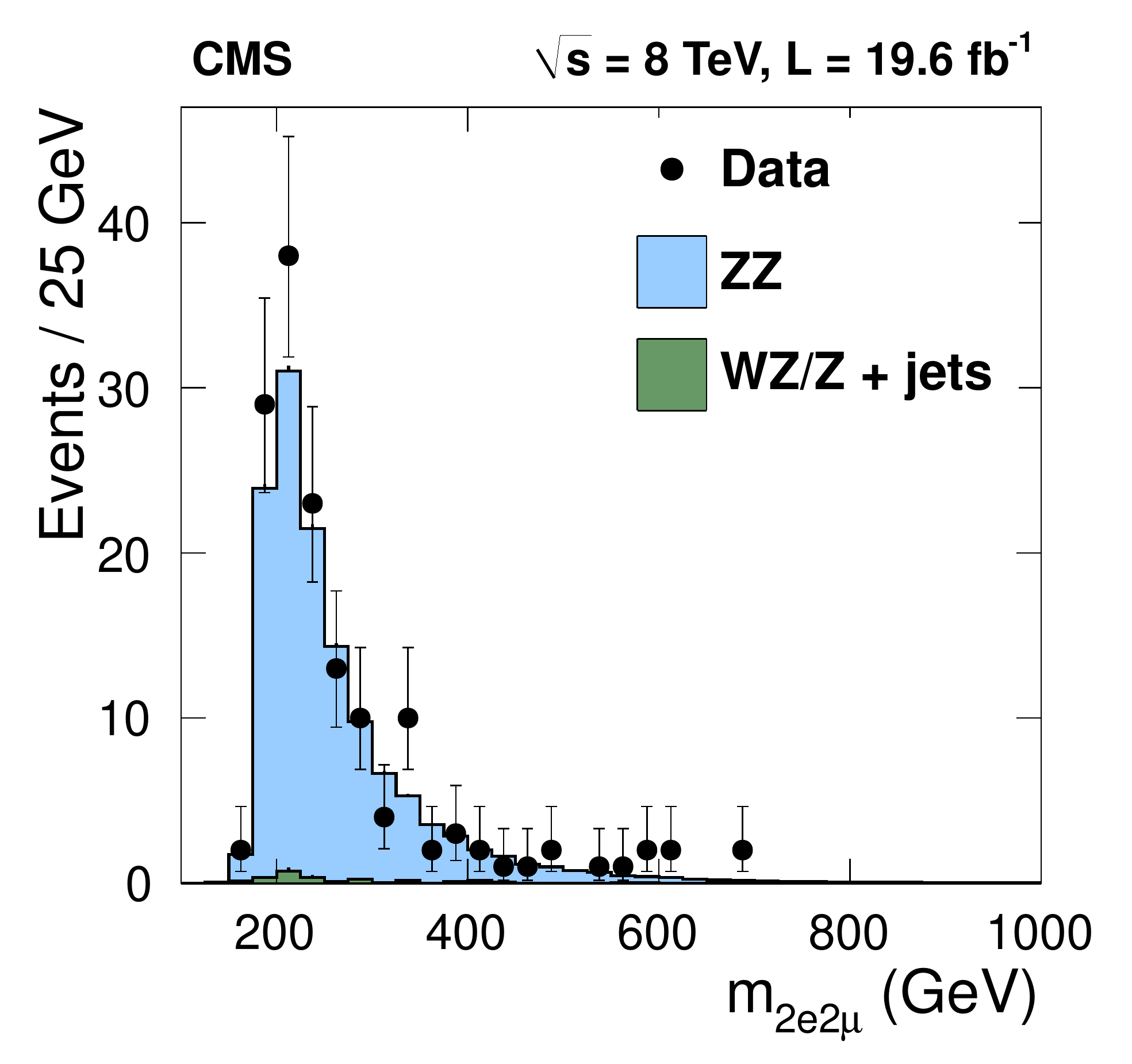}
\includegraphics[width=0.45\textwidth]{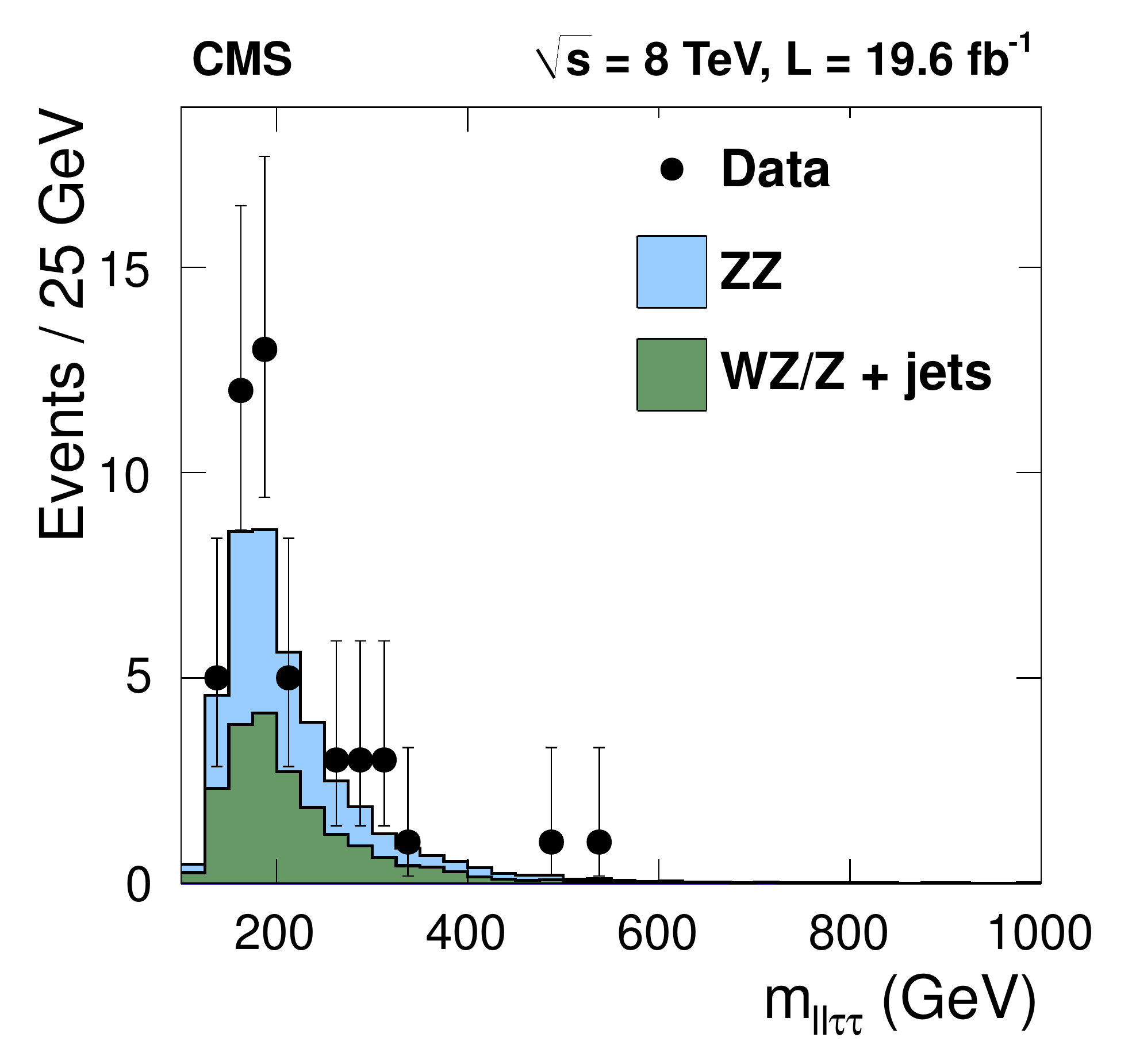}
   \caption{
Distribution of the reconstructed four-lepton mass for the (upper left)
          $4\Pe$, (upper right) $4\Pgm$, (lower left) $2\Pe2\Pgm$, and (lower right) combined $\taus$ decay channels.
The data sample corresponds to an integrated luminosity of 19.6\fbinv.
Points represent the data, the shaded histograms labeled $\cPZ\cPZ$
represent
the \POWHEG+{\textsc{gg2zz}}+\PYTHIA predictions
for $\PZ\PZ$ signal, the histograms labeled $\PW\cPZ/\cPZ$+jets
show the background, which is
estimated from data,
as
described in the text.}
   \label{fig:MassDistribution}
\end{figure*}

\begin{figure*}[htbp]
\centering
\includegraphics[width=0.45\textwidth]{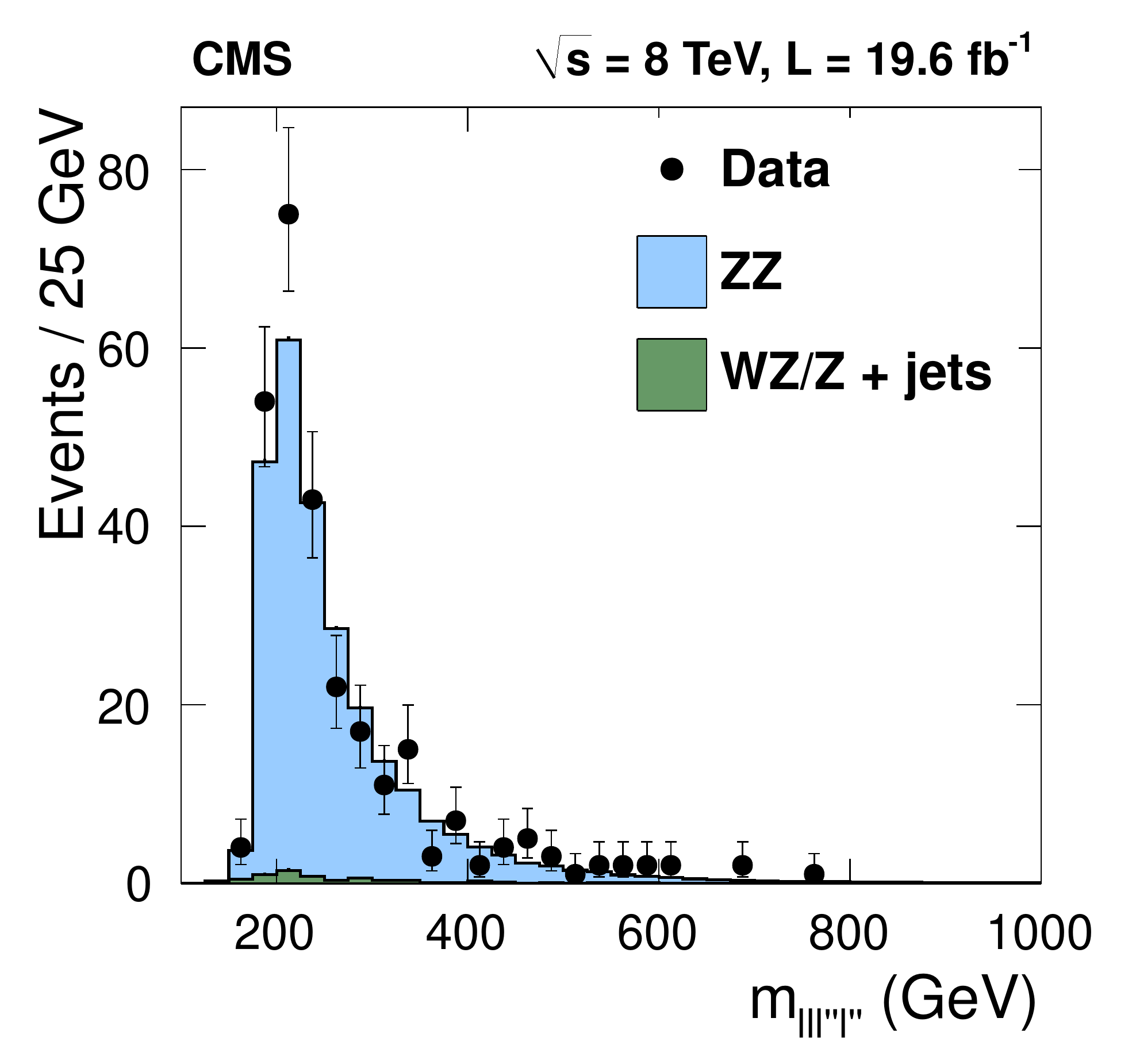}
\includegraphics[width=0.45\textwidth]{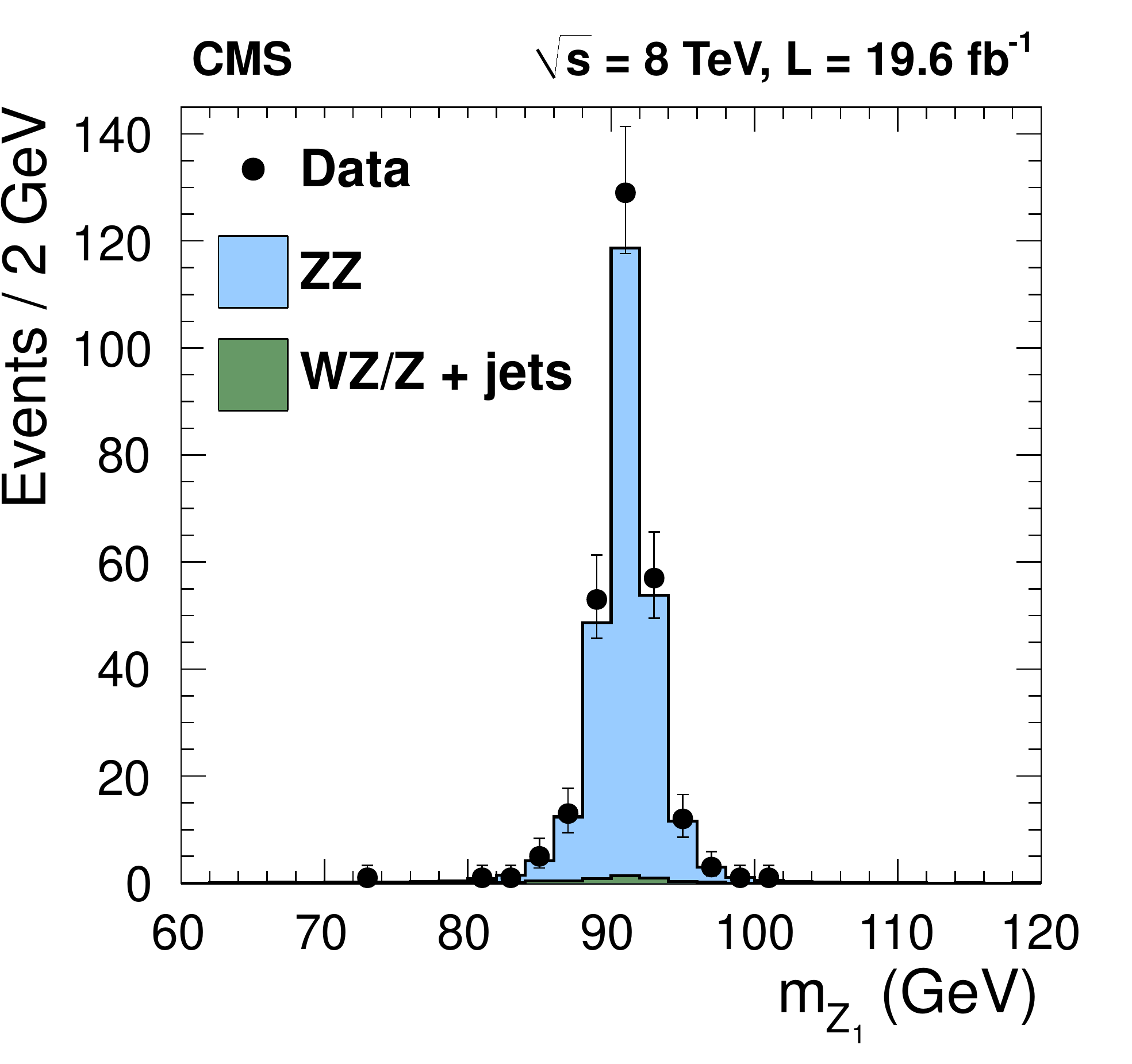}
\includegraphics[width=0.45\textwidth]{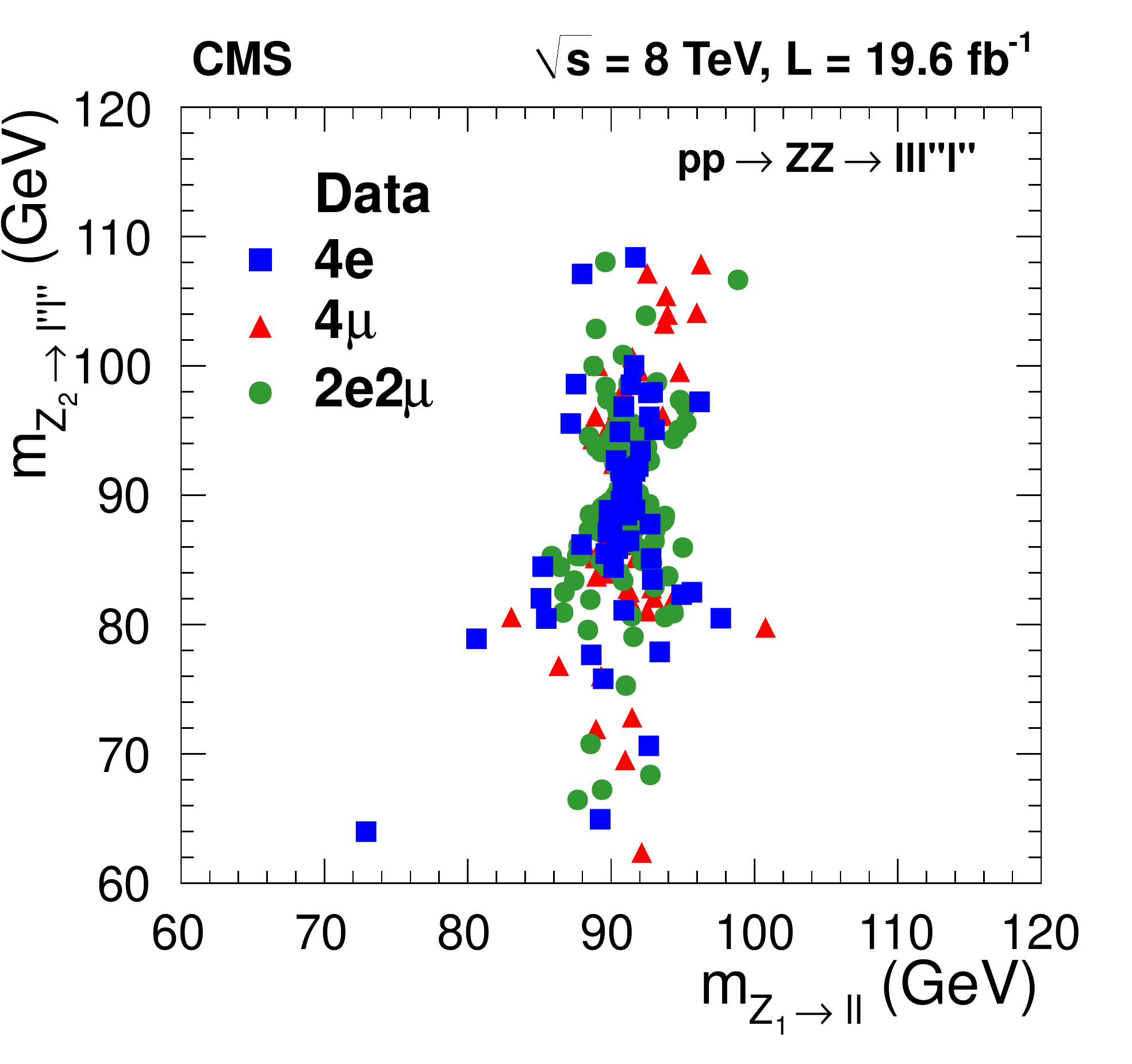}
\includegraphics[width=0.45\textwidth]{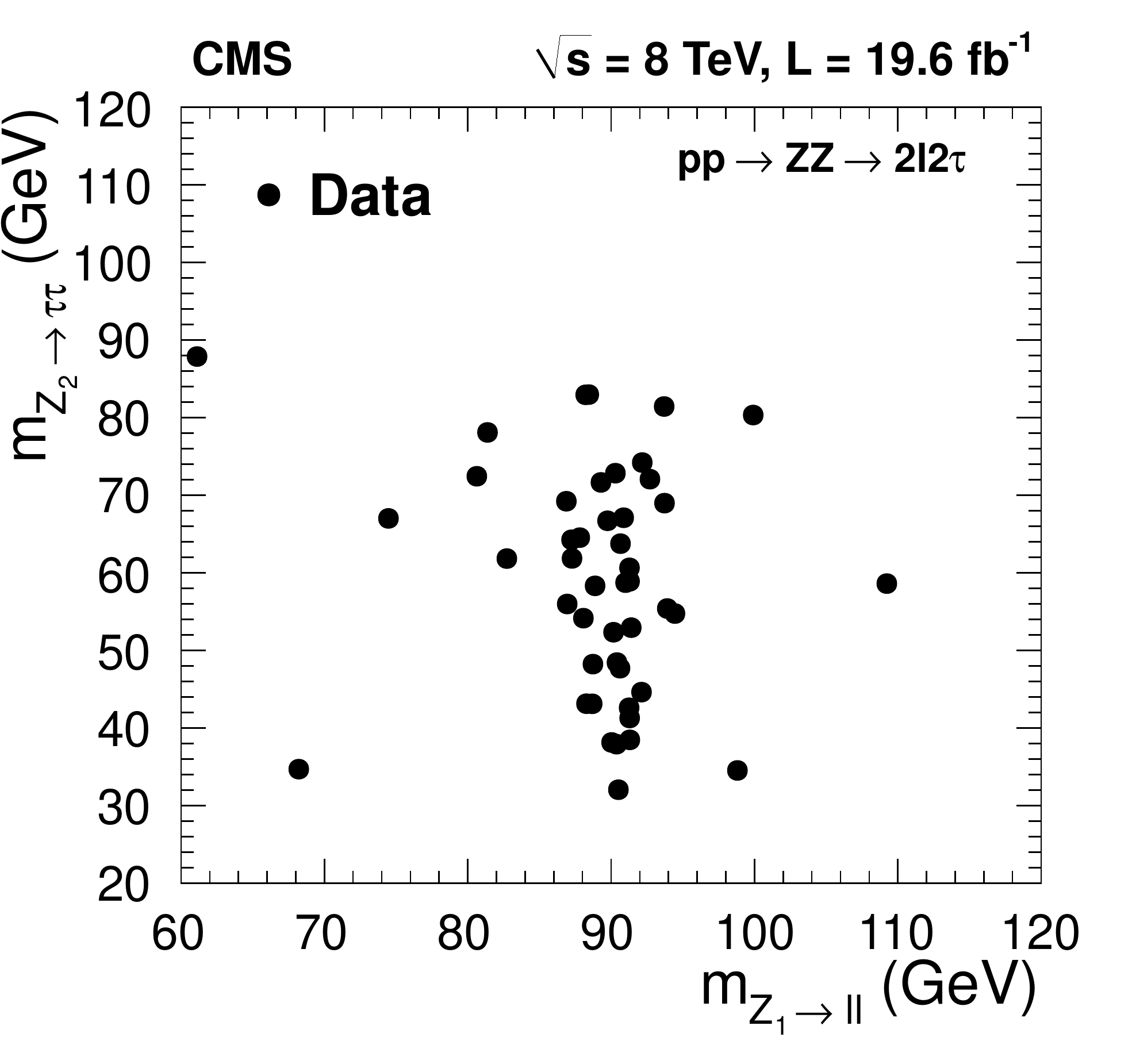}
\caption{(upper left) Distribution of the reconstructed
four-lepton  mass for the sum of the
$4\Pe$, $4\Pgm$, and $2\Pe2\Pgm$ decay channels.
(upper right) Reconstructed $\cPZ_1$ mass.
 The correlation between
the reconstructed $\PZ_1$ and $\PZ_2$ masses for
the (lower left) combined $4\Pe$, $4\Pgm$, and $2\Pe2\Pgm$ final states and (lower right) for $\taus$
final states.
Points represent the data, the shaded histograms labeled $\cPZ\cPZ$
represent
the \POWHEG+{\textsc{gg2zz}}+\PYTHIA predictions
for $\PZ\PZ$ signal, the histograms labeled $\PW\cPZ/\cPZ$+jets
show  background, which is
estimated from data,
as
described in the text.}
\label{fig:Mass4l}
\end{figure*}

The measured and expected event yields for all decay channels
are summarized
in Table~\ref{table:results}.
The recently discovered Higgs particle with the mass of 125\GeV does not contribute to this analysis as background because of the phase space selection requirements.
The reconstructed four-lepton invariant mass distributions for the $4\Pe$, $4\Pgm$, $2\Pe2\Pgm$,
and combined $\taus$
decay  channels
are shown in Fig.~\ref{fig:MassDistribution}.
The shape of the background is taken from data.
The reconstructed four-lepton invariant mass distribution for
the combined $4\Pe$, $4\Pgm$, and $2\Pe2\Pgm$ channels is shown in Fig.~\ref{fig:Mass4l}(upper left). Figure~\ref{fig:Mass4l}(upper right)
presents the invariant mass of the $\cPZ_1$ candidates. Figures~\ref{fig:Mass4l}(lower left) and (lower right)
show the correlation between the reconstructed $\cPZ_1$ and $\cPZ_2$ masses for
(lower left) $4\Pe$, $4\Pgm$, and $2\Pe2\Pgm$ and for (lower right) $\taus$ final states.
The data are well reproduced by the signal
simulation and with background predictions estimated from data.

\begin{table*}[htb]
\centering
\topcaption{ The expected yields of $\PZ\PZ$ and
 background
events, as well as their sum
(``Total expected'') are
compared with the observed yields for each decay channel. The statistical and systematic uncertainties are also shown.}
\begin{tabular}{ccccc}
\hline
Decay & Expected  &  Background   & Total & Observed \\
channel & $\cPZ\cPZ$ yield  &  & expected  & \\
\hline
$4\Pe$ 			& $ 55.28 \pm 0.25 \pm 7.64 $ & $ 2.16 \pm 0.26 \pm 0.88 $ & $57.44 \pm 0.37 \pm 7.69 $ & $  54 $\\
$4\Pgm$ 		& $ 77.32 \pm 0.29 \pm 10.08 $ & $ 1.19 \pm 0.36 \pm 0.48 $ & $78.51 \pm 0.49 \pm 10.09 $ & $ 75 $\\
$2\Pe 2\Pgm$ 		& $ 136.09 \pm 0.59 \pm 17.50 $ & $ 2.35 \pm 0.34 \pm 0.93 $ & $138.44 \pm 0.70 \pm 17.52 $ & $ 148 $\\
\hline
$\Pe\Pe\tauh\tauh$      &$2.46\pm 0.03 \pm 0.32$ & $ 3.46 \pm 0.34 \pm 1.04 $ & $ 5.92 \pm 0.36 \pm 1.15 $ & 10 \\
$\Pgm\Pgm\tauh\tauh$  & $2.80 \pm 0.03 \pm 0.34$  & $ 3.89 \pm 0.37 \pm 1.17 $ & $ 6.69 \pm 0.39 \pm 1.30 $ & 10 \\
$\Pe\Pe\Pgt_{\Pe}\tauh$      & $2.79 \pm 0.03 \pm 0.36$  & $ 3.87 \pm 1.26 \pm 1.16 $ & $ 6.66 \pm 1.34 \pm 1.29 $ & 9 \\
$\Pgm\Pgm\Pgt_{\Pe}\tauh$  & $2.87 \pm 0.03 \pm 0.37$  & $ 1.49 \pm 0.67 \pm 0.60 $ & $ 4.36 \pm 0.71 \pm 0.73 $ & 2 \\
$\Pe\Pe\Pgt_{\Pgm}\tauh$    & $ 3.27 \pm 0.03 \pm 0.42$  & $ 1.47 \pm 0.41 \pm 0.44  $ & $ 4.74 \pm 0.43 \pm 0.63 $ & 2 \\
$\Pgm\Pgm\Pgt_{\Pgm}\tauh$  & $ 3.81 \pm 0.03 \pm 0.50$  & $ 1.55 \pm 0.43  \pm 0.46$ & $ 5.36 \pm 0.46 \pm 0.70  $  & 5 \\
$\Pe\Pe\Pgt_{\Pe}\Pgt_{\Pgm}$    & $ 2.23 \pm 0.03 \pm 0.29$  & $ 3.04 \pm 1.32 \pm 1.50 $ & $ 5.27 \pm 1.40 \pm 1.61 $   & 4 \\
$\Pgm\Pgm\Pgt_{\Pe}\Pgt_{\Pgm}$  &$2.41\pm 0.03 \pm 0.32$ & $ 0.74 \pm 0.51 \pm 0.37 $ & $ 3.15 \pm 0.54 \pm 0.51 $ & 5 \\
\hline
Total $\taus$                   & $22.65 \pm 0.05 \pm 2.94$ & $ 19.51 \pm 2.15 \pm 5.85  $ & $ 42.16 \pm 2.28 \pm 6.87 $  & 47 \\
\hline
\end{tabular}
\label{table:results}
\end{table*}

The measured yields are used to evaluate the total $\cPZ\cPZ$
production cross section.
The signal acceptance is evaluated from simulation and corrected
for each individual lepton flavor in bins of $\pt$ and $\eta$
using factors obtained with
the  ``tag-and-probe'' technique.
The requirements on $\pt$ and $\eta$ for the particles in the final state reduce the full possible
phase space of the $\ZZ \to 4\ell$ measurement by a factor within a range of 0.56--0.59 for the
4$\Pe$, 4$\Pgm$, and 2$\Pe$2$\Pgm$, depending on the final state, and
by a factor of 0.18--0.21 for the $\taus$ final states, with respect to all events generated
in the mass window $60 < m_{\cPZ_1},m_{\cPZ_2} < 120\GeV$.
The branching fraction for $\PZ \to \ell'\ell'$ is $(3.3658 \pm 0.0023)\%$
for each lepton flavor~\cite{Beringer:1900zz}.

To include all final states in the cross section calculation,
a simultaneous fit to the number of  observed events in all decay channels
is performed.
The likelihood is written as a combination of
individual channel likelihoods for
the signal and background hypotheses,
with 
systematical uncertainties used as nuisance parameters in the fit.
Each $\Pgt$-lepton decay mode, listed in Table~\ref{table:results}, is treated as a separate channel.

Table~\ref{table:crosssections}
lists the total cross section obtained from each
individual decay channel as well
as the total cross
section based on the combination of all channels.
The measured cross section agrees with the theoretical value of
$7.7 \pm 0.6\unit{pb}$
 calculated with {\MCFM} 6.0. In this calculation, the contribution 
from 
$ \Pq\Paq \to \cPZ\cPZ$ is obtained at NLO, while the smaller
contribution (approximately 6\%) from
$\Pg\Pg \to \cPZ\cPZ$ is obtained at LO. 
The MSTW2008 PDF is used
and the renormalization and factorization scales set to $\mu_{\mathrm{R}} = \mu_{\mathrm{F}} = 91.2\GeV$. 

\begin{table*}[htb]
\centering
\topcaption{ The total $\cPZ\cPZ$ production cross section as measured
in each decay channel
and for the combination of all channels.}
\renewcommand{\arraystretch}{1.2}
\begin{tabular}{cc}
\hline
Decay channel & Total cross section, pb \\
\hline
$4\Pe$ 			& $7.2\, ^{+1.0}_{-0.9}\stat\, ^{+0.6} _{-0.5}\syst \pm 0.4\theo \pm 0.2\lum$ \\
$4\Pgm$ 		& $7.3\, ^{+0.8}_{-0.8}\stat\, ^{+0.6} _{-0.5}\syst \pm 0.4\theo \pm 0.2\lum$ \\
$2\Pe 2\Pgm$ 		& $8.1\, ^{+0.7}_{-0.6}\stat\, ^{+0.6} _{-0.5}\syst \pm 0.4\theo \pm 0.2\lum$ \\
$\ell\ell\Pgt\Pgt$ 		& $7.7\, ^{+2.1}_{-1.9}\stat\, ^{+2.0} _{-1.8}\syst \pm 0.4\theo \pm 0.2\lum$ \\
\hline
Combined                           & $7.7\, \pm 0.5\stat\, ^{+0.5} _{-0.4}\syst \pm 0.4\theo \pm 0.2\lum$ \\
\hline
\end{tabular}
\label{table:crosssections}
\end{table*}

The measurement of the differential cross sections is an important part of this analysis, since it
provides detailed information about $\cPZ\cPZ$ kinematics. Three decay channels, $4\Pe$, $4\Pgm$,  and $2\Pe 2\Pgm$, are
combined, since their kinematic distributions are the same;
the $\taus$ channel is not included.
The observed yields are unfolded using the
method described in Ref.~\cite{unfolding}.

\begin{figure*}[htbp]
\centering
\includegraphics[width=0.45\textwidth]{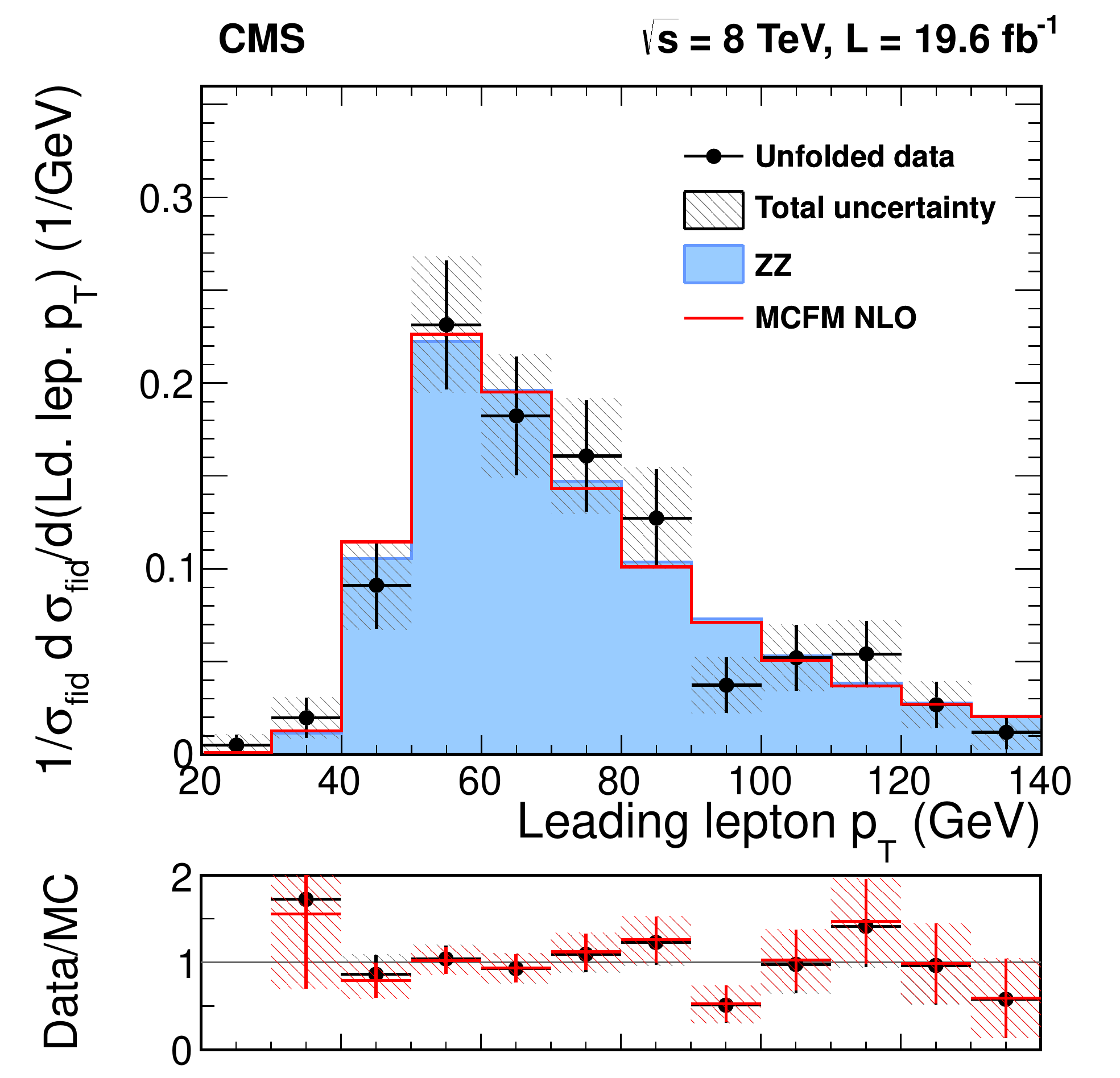} \includegraphics[width=0.45\textwidth]{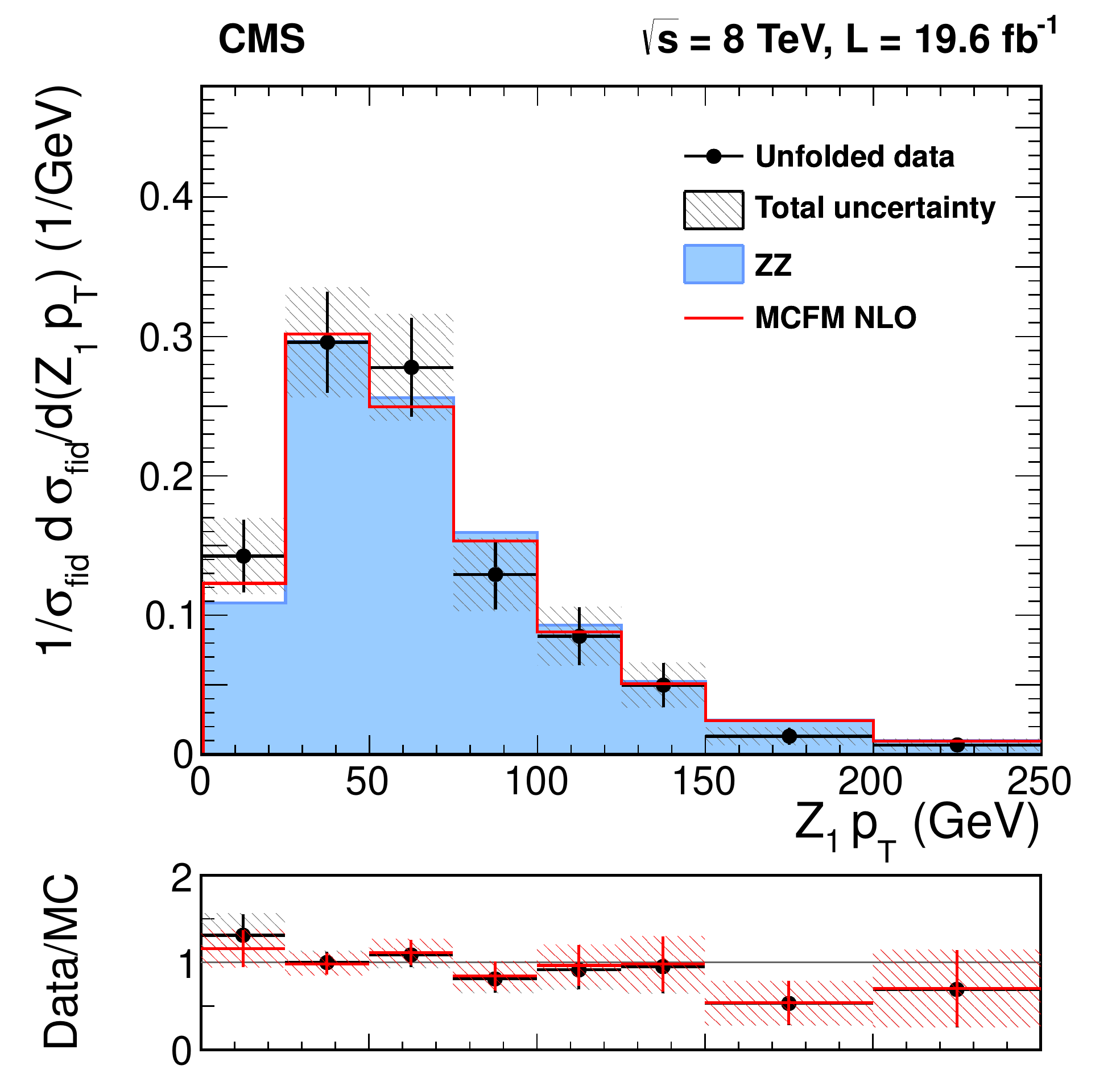}
\includegraphics[width=0.45\textwidth]{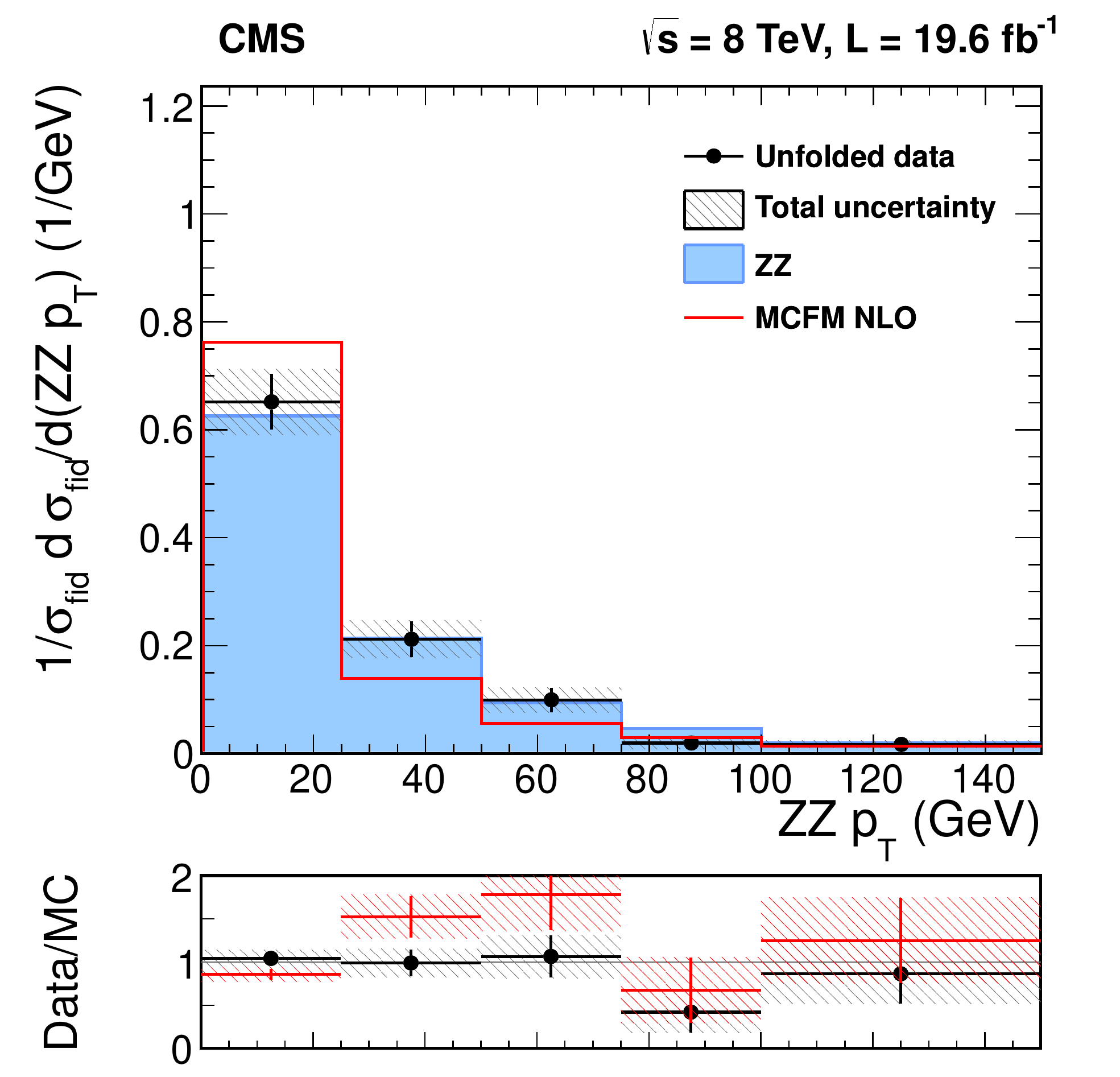}
\includegraphics[width=0.45\textwidth]{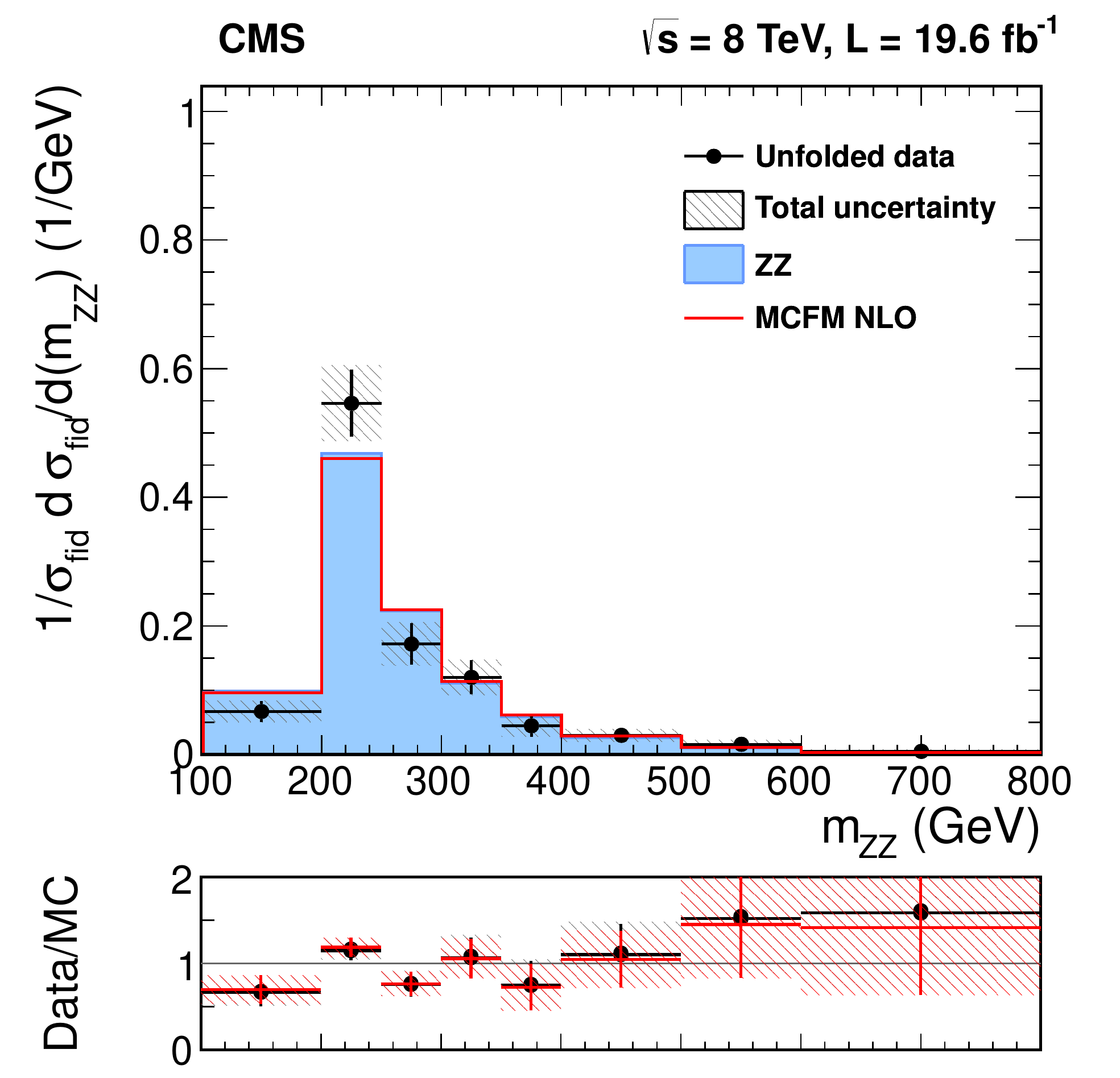} \caption{Differential cross sections normalized to the fiducial cross section
for the combined 4$\Pe$, 4$\Pgm$, and 2$\Pe$2$\Pgm$ decay channels as a function of
$\pt$ for (upper left) the highest \pt lepton in the
event, (upper right) the $\cPZ_1$, and (lower left) the $\cPZ\cPZ$ system. Figure (lower right) shows the normalized
$\rd\sigma/\rd{m_{\cPZ\cPZ}}$
distribution.
Points represent the data, and the shaded histograms labeled $\cPZ\cPZ$
represent
the \POWHEG+{\textsc{gg2zz}}+\PYTHIA predictions
for $\PZ\PZ$ signal,
while the solid curves correspond to results of the {\MCFM} calculations.
The bottom part of each subfigure represents the ratio of the measured cross section
to the expected one from \POWHEG+{\textsc{gg2zz}}+\PYTHIA (black crosses with solid symbols) and {\MCFM } (red crosses). The shaded
areas on all the plots represent the full uncertainties calculated as the quadrature sum
of the statistical and systematic uncertainties, whereas the crosses represent the statistical
uncertainties only.
}
\label{fig:diff1}
\end{figure*}

\begin{figure}[htbp]
\centering
\includegraphics[width=0.45\textwidth]{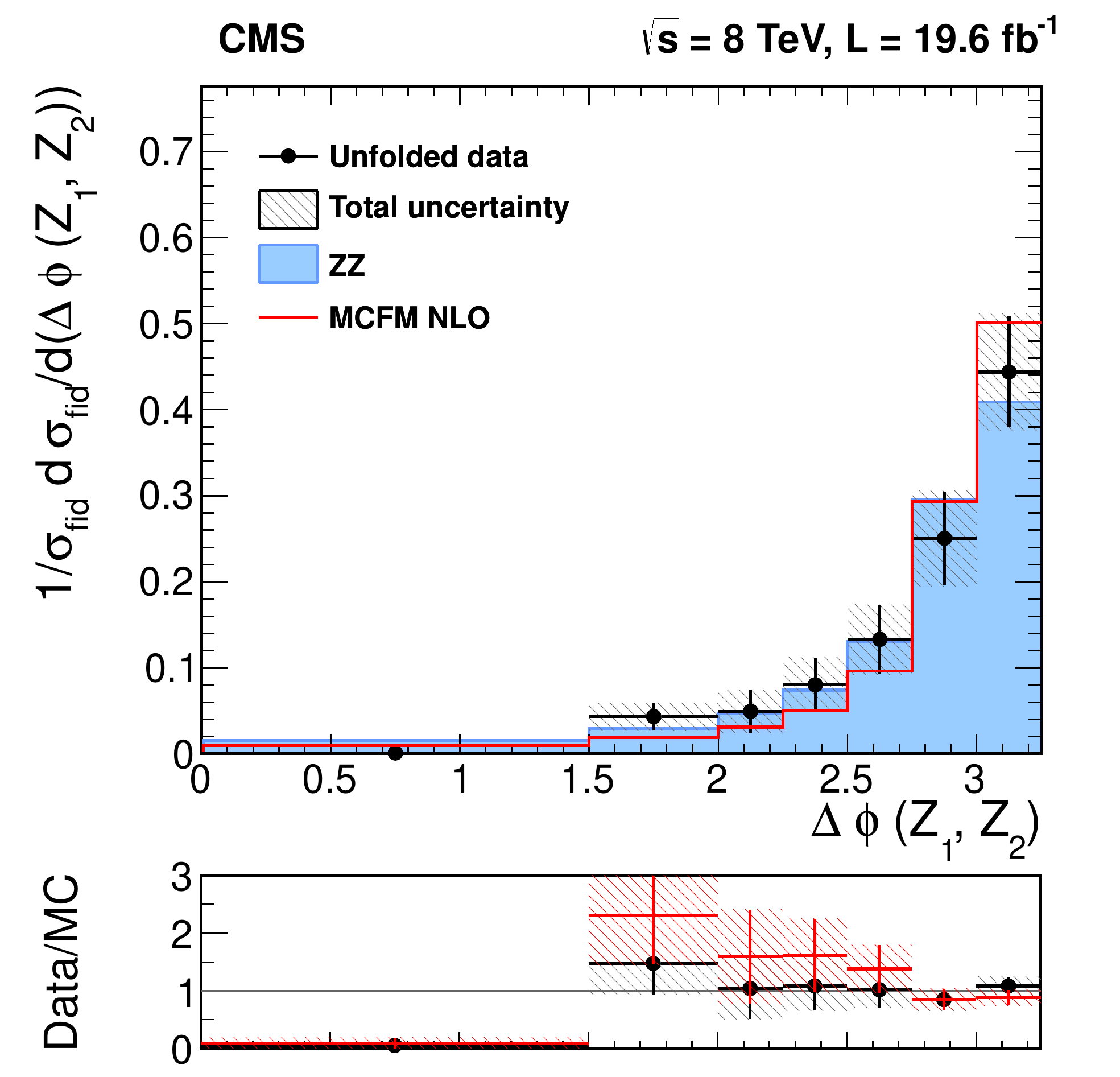} \includegraphics[width=0.45\textwidth]{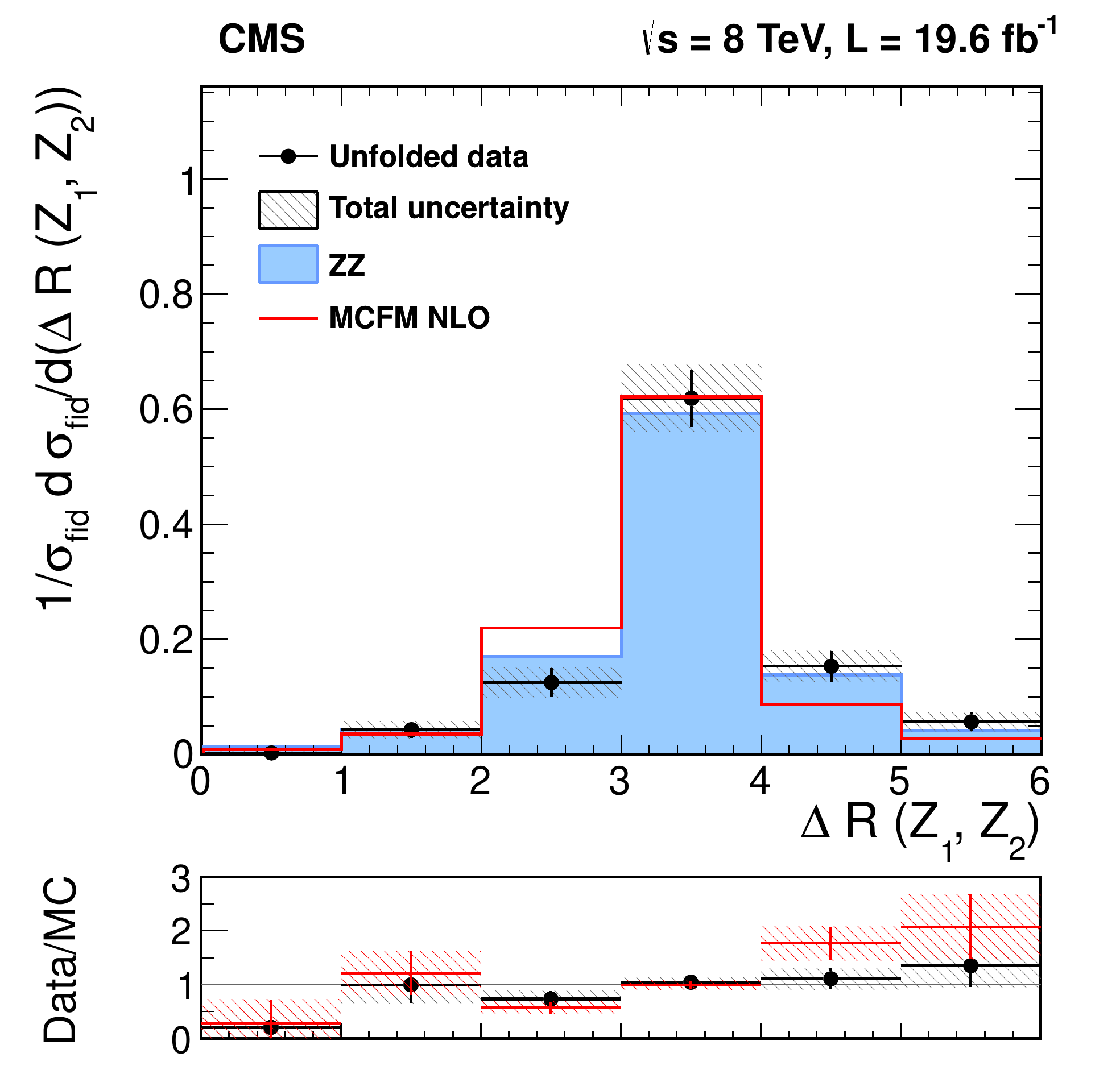}
\caption{Differential cross section normalized to the fiducial cross section
for the combined 4$\Pe$, 4$\Pgm$, and 2$\Pe$2$\Pgm$ decay channels as a function
of (\cmsLeft) azimuthal separation of the two $\cPZ$ bosons and
(\cmsRight) $\Delta R$ between the $\cPZ$-bosons.
Points represent the data, and the shaded histograms labeled $\cPZ\cPZ$
represent
the \POWHEG+{\textsc{gg2zz}}+\PYTHIA predictions
for $\PZ\PZ$ signal,
while the solid curves correspond to results of the \MCFM calculations.
The bottom part of each subfigure represents the ratio of the measured cross section
to the expected one from \POWHEG+{\textsc{gg2zz}}+\PYTHIA (black crosses with solid symbols) and \MCFM (red crosses). The shaded
areas on all the plots represent the full uncertainties calculated as the quadrature sum
of the statistical and systematic uncertainties, whereas the crosses represent the statistical
uncertainties only.
}
\label{fig:diff2}
\end{figure}

The differential distributions normalized to the fiducial cross sections are
presented
in Figs.~\ref{fig:diff1} and \ref{fig:diff2} for the combination of the 4$\Pe$, 4$\Pgm$, and 2$\Pe$2$\Pgm$
decay channels. The fiducial cross section definition includes $\pt^{\ell}$ and
$\abs{\eta^{\ell}}$ selections on each lepton,
and the 60--120\GeV mass requirement, as described in Section 4.
Figure~\ref{fig:diff1} shows the differential cross sections in bins of $\pt$ for:
(upper left) the highest-\pt lepton in the
event, (upper right) the $\cPZ_1$, and (lower left) the $\cPZ\cPZ$ system. Figure~\ref{fig:diff1}(lower left) shows the normalized
$\rd\sigma/\rd{m_{\cPZ\cPZ}}$
distribution. The data are corrected for background contributions and compared with
the theoretical predictions from \POWHEG and \MCFM. The bottom part of each plot shows
the ratio of the measured to the predicted values. The bin sizes were chosen according to the resolution of the relevant variables, trying also to keep the statistical
uncertainties at a similar level for all the bins.
Figure~\ref{fig:diff2} shows the angular correlations between
$\cPZ$ bosons, which are in good agreement with the MC simulations.
Some difference between
\POWHEG and \MCFM calculations appears at very low $\pt$ of the $\ZZ$ system
and  for azimuthal separation of the $\cPZ$
bosons close to $\pi$. This region
is better modeled by \POWHEG interfaced with the \PYTHIA
parton shower program.

\section{Limits on anomalous triple gauge couplings}

\begin{figure}[htbp]
\centering
\includegraphics[width=\cmsFigWidth]{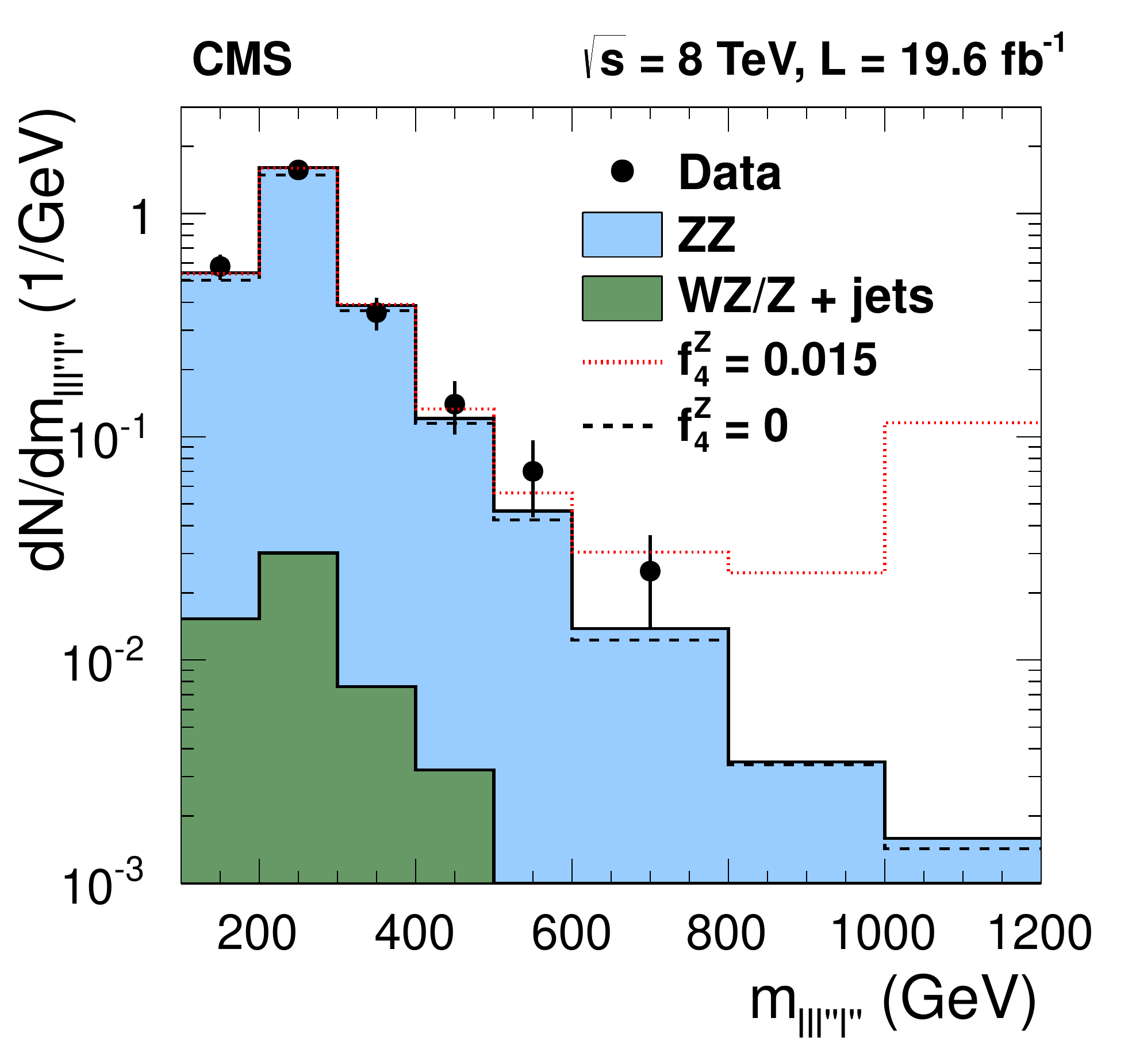}
\caption{Distribution of the four-lepton reconstructed mass for the combined
$4\Pe$, 4$\Pgm$, and $2\Pe2\Pgm$ channels.
Points represent the data, the shaded histogram labeled $\cPZ\cPZ$
represents
the \POWHEG+{\textsc{gg2zz}}+\PYTHIA predictions
for $\PZ\PZ$ signal, the histograms labeled $\PW\cPZ/\cPZ$+jets
shows  background, which is
estimated from data, as described in the text.
The dashed and dotted histograms indicate the results of the \SHERPA
simulation for the SM ($f_4^{\cPZ} = 0$) and
 in the presence of an ATGC ($f_4^{\cPZ}=0.015$) with
all the other anomalous couplings set to zero.
The last bin includes all entries with masses above 1000\GeV.
 }
\label{figure:sherpa4l}
\end{figure}

The presence of ATGCs
would be manifested as an increased yield of events at high four-lepton masses.
Figure~\ref{figure:sherpa4l} presents the
distribution of the four-lepton reconstructed mass,
which is used to set the limits,
for the combined
$4\Pe$, 4$\Pgm$, and $2\Pe2\Pgm$ channels.
The shaded histogram represents the results of the \POWHEG
simulation for the $\cPZ\cPZ$ signal, and the dashed line, which agrees well with it, is the prediction of
\SHERPA for $f_4^{\cPZ} = 0$ normalized to the \MCFM cross section.
The dotted line indicates the \SHERPA predictions
for a specific ATGC value ($f_4^{\cPZ}=0.015$) with
all the other anomalous couplings set to zero.

\begin{figure}[htbp]
\centering
\includegraphics[width=0.48\textwidth]{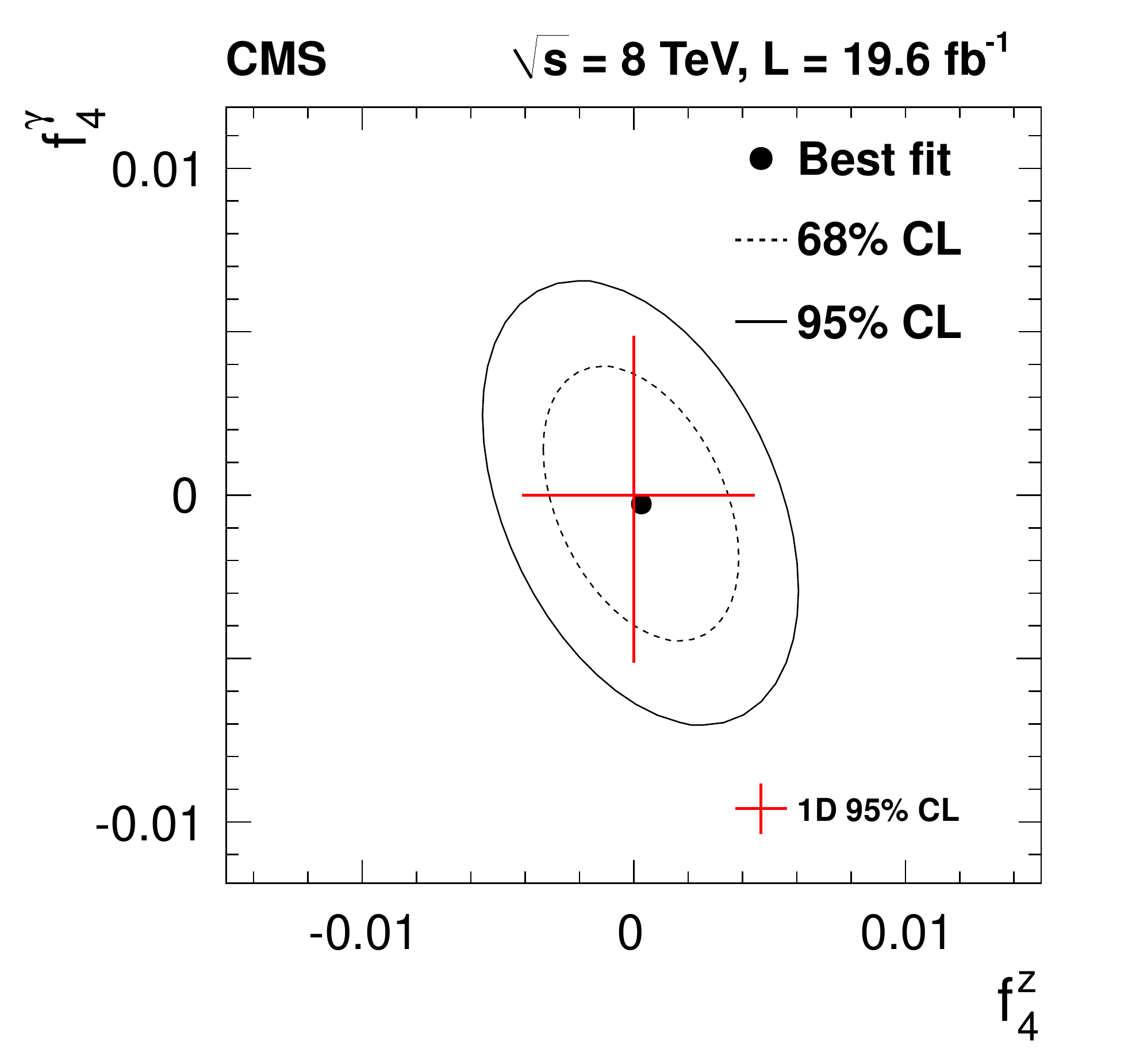}
\includegraphics[width=0.48\textwidth]{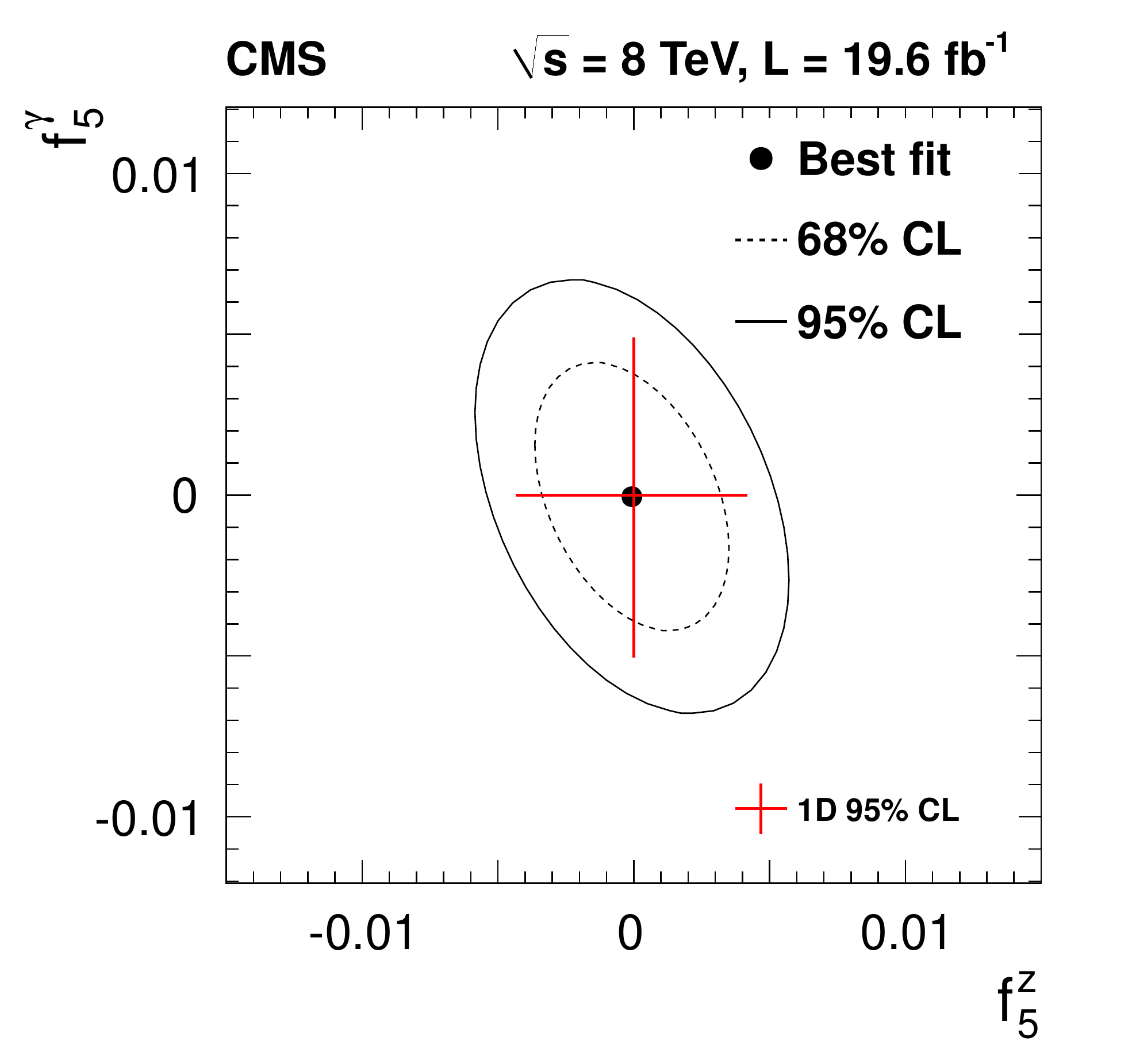}
\caption{Two-dimensional exclusion limits at 68\% (dashed contour) and 95\% (solid contour) CL on the $\PZ\PZ\PZ$ and $\PZ\PZ\gamma$ ATGCs. The \cmsLeft (\cmsRight) plot shows the exclusion contour in the $(f_{4(5)}^\cPZ, f_{4(5)}^\gamma)$ parameter planes. The solid dot shows where the likelihood reaches its maximum. The values of couplings outside of contours are excluded at the corresponding confidence level. The lines in the middle represent one-dimensional limits. No form factor is used. }
\label{figure:aTGC}
\end{figure}

The invariant mass distributions are interpolated from the
\SHERPA simulation for different values of the anomalous couplings in the range between 0 and 0.015. For
each distribution, only one or two couplings are varied while all others are set to zero.
The measured signal is obtained from a comparison of the data to
a grid of ATGC models in the
$(f_{4}^\cPZ, f_{4}^\gamma)$ and
$(f_{5}^\cPZ, f_{5}^\gamma)$
parameter
planes.  Expected signal values are interpolated between the 2D grid points using a second-degree polynomial,
 since the cross
section for signal depends quadratically on the coupling parameters.
A profile likelihood method~\cite{Beringer:1900zz} is used to derive the limits. Systematic uncertainties are taken into account by
varying the number of expected signal and background events within their uncertainties.
No form factor is used when deriving the limits so that the
results do not depend  on any assumed energy scale characterizing new physics.
The constraints on anomalous couplings are displayed in Fig.~\ref{figure:aTGC}.
The curves indicate
68\% and 95\% confidence levels, and
the solid dot shows where the likelihood reaches
its maximum. Coupling values outside the contours
are excluded at the corresponding confidence levels.
The limits are dominated by statistical uncertainties.

One-dimensional 95\% CL limits
for the $f_4^{\cPZ,\gamma}$ and $f_5^{\cPZ,\gamma}$ anomalous coupling parameters are:
\ifthenelse{\boolean{cms@external}}{
\begin{align*}
-0.004&<f_4^\cPZ<0.004,\\
-0.004&<f_5^\cPZ<0.004,\\
-0.005&<f_4^{\gamma}<0.005,\\
-0.005&<f_5^{\gamma}<0.005.
\end{align*}
}{
\begin{equation*}
-0.004<f_4^\cPZ<0.004,\quad -0.004<f_5^\cPZ<0.004,\quad -0.005<f_4^{\gamma}<0.005,\quad -0.005<f_5^{\gamma}<0.005.
\end{equation*}
}
In the one-dimensional fits, all of the ATGC
parameters except the one under study are set to zero.
These values extend previous CMS
results on vector boson self-interactions~\cite{SMP-12-007} and improve on the previous limits by factors of three to four, they are presented in Fig.~\ref{figure:aTGC} as horizontal and vertical lines.

\section{Summary}

Measurements have been presented of the inclusive $\cPZ\cPZ$ production
cross section
in proton-proton collisions at 8\TeV in the $\cPZ \cPZ \to \ell\ell\ell'\ell'$ decay mode,
with $\ell = \Pe, \Pgm$ and $\ell' = \Pe, \Pgm, \Pgt$.
The data sample corresponds to an integrated luminosity of $19.6\fbinv$.
The measured total cross section
$\sigma ( \pp \to \PZ\PZ) = 7.7\, \pm 0.5\stat\, ^{+0.5} _{-0.4}\syst \pm 0.4\theo \pm 0.2\lum\unit{pb}$ 
for both $\cPZ$ bosons in the mass range $60 < m_\cPZ < 120\GeV$
and the differential cross sections agree well with
the SM predictions.
Improved limits on
anomalous $\PZ \PZ \PZ$ and $\PZ \PZ \gamma$ triple gauge
couplings
are established,
significantly restricting their possible allowed ranges.

\section*{Acknowledgements}

We congratulate our colleagues in the CERN accelerator departments for the excellent performance of the LHC and thank the technical and administrative staffs at CERN and at other CMS institutes for their contributions to the success of the CMS effort. In addition, we gratefully acknowledge the computing centres and personnel of the Worldwide LHC Computing Grid for delivering so effectively the computing infrastructure essential to our analyses. Finally, we acknowledge the enduring support for the construction and operation of the LHC and the CMS detector provided by the following funding agencies: BMWFW and FWF (Austria); FNRS and FWO (Belgium); CNPq, CAPES, FAPERJ, and FAPESP (Brazil); MES (Bulgaria); CERN; CAS, MoST, and NSFC (China); COLCIENCIAS (Colombia); MSES and CSF (Croatia); RPF (Cyprus); MoER, ERC IUT and ERDF (Estonia); Academy of Finland, MEC, and HIP (Finland); CEA and CNRS/IN2P3 (France); BMBF, DFG, and HGF (Germany); GSRT (Greece); OTKA and NIH (Hungary); DAE and DST (India); IPM (Iran); SFI (Ireland); INFN (Italy); NRF and WCU (Republic of Korea); LAS (Lithuania); MOE and UM (Malaysia); CINVESTAV, CONACYT, SEP, and UASLP-FAI (Mexico); MBIE (New Zealand); PAEC (Pakistan); MSHE and NSC (Poland); FCT (Portugal); JINR (Dubna); MON, RosAtom, RAS and RFBR (Russia); MESTD (Serbia); SEIDI and CPAN (Spain); Swiss Funding Agencies (Switzerland); MST (Taipei); ThEPCenter, IPST, STAR and NSTDA (Thailand); TUBITAK and TAEK (Turkey); NASU and SFFR (Ukraine); STFC (United Kingdom); DOE and NSF (USA).

Individuals have received support from the Marie-Curie programme and the European Research Council and EPLANET (European Union); the Leventis Foundation; the A. P. Sloan Foundation; the Alexander von Humboldt Foundation; the Belgian Federal Science Policy Office; the Fonds pour la Formation \`a la Recherche dans l'Industrie et dans l'Agriculture (FRIA-Belgium); the Agentschap voor Innovatie door Wetenschap en Technologie (IWT-Belgium); the Ministry of Education, Youth and Sports (MEYS) of the Czech Republic; the Council of Science and Industrial Research, India; the HOMING PLUS programme of Foundation for Polish Science, cofinanced from European Union, Regional Development Fund; the Compagnia di San Paolo (Torino); and the Thalis and Aristeia programmes cofinanced by EU-ESF and the Greek NSRF.

\bibliography{auto_generated}   

\cleardoublepage \appendix\section{The CMS Collaboration \label{app:collab}}\begin{sloppypar}\hyphenpenalty=5000\widowpenalty=500\clubpenalty=5000\textbf{Yerevan Physics Institute,  Yerevan,  Armenia}\\*[0pt]
V.~Khachatryan, A.M.~Sirunyan, A.~Tumasyan
\vskip\cmsinstskip
\textbf{Institut f\"{u}r Hochenergiephysik der OeAW,  Wien,  Austria}\\*[0pt]
W.~Adam, T.~Bergauer, M.~Dragicevic, J.~Er\"{o}, C.~Fabjan\cmsAuthorMark{1}, M.~Friedl, R.~Fr\"{u}hwirth\cmsAuthorMark{1}, V.M.~Ghete, C.~Hartl, N.~H\"{o}rmann, J.~Hrubec, M.~Jeitler\cmsAuthorMark{1}, W.~Kiesenhofer, V.~Kn\"{u}nz, M.~Krammer\cmsAuthorMark{1}, I.~Kr\"{a}tschmer, D.~Liko, I.~Mikulec, D.~Rabady\cmsAuthorMark{2}, B.~Rahbaran, H.~Rohringer, R.~Sch\"{o}fbeck, J.~Strauss, A.~Taurok, W.~Treberer-Treberspurg, W.~Waltenberger, C.-E.~Wulz\cmsAuthorMark{1}
% ...
\vskip\cmsinstskip
% ...
1:~~Also at Vienna University of Technology, Vienna, Austria\\
2:~~Also at CERN, European Organization for Nuclear Research, Geneva, Switzerland\\
% ...

\end{sloppypar}
\end{document}